\documentclass{article}

\input{style/preamble.sty}

\begin{document}
%
%
%
%

\begin{titlepage}
	\title{More Efficient Exact Group-Invariance Testing: using a Representative Subgroup}
	\date{\today}
	
%
%
%
%

\author{Nick W. Koning\footnote{Econometric Institute, Erasmus University Rotterdam, P.O. Box 1738,
3000 DR Rotterdam, The Netherlands. e-mail: n.w.koning@ese.eur.nl.}{\ \ }and Jesse Hemerik\footnote{Department Biometris, Wageningen University \& Research, P.O. Box 16, 6700 AC Wageningen, The Netherlands. e-mail: jesse.hemerik@wur.nl.}}

	
	\maketitle
	

\begin{abstract}
	Non-parametric tests based on permutation, rotation or sign-flipping are examples of group-invariance tests.
	These tests test invariance of the null distribution under a set of transformations that has a group structure, in the algebraic sense.
	Such groups are often huge, which makes it computationally infeasible to test using the entire group.
	Hence, it is standard practice to test using a randomly sampled set of transformations from the group.
	This random sample still needs to be substantial to obtain good power and replicability.
	We improve upon this standard practice by using a well-designed subgroup of transformations instead of a random sample.
	The resulting subgroup-invariance test is still exact, as invariance under a group implies invariance under its subgroups.
	
	We illustrate this in a generalized location model and obtain more powerful tests based on the same number of transformations.
	In particular, we show that a subgroup-invariance test is consistent for lower signal-to-noise ratios than a test based on a random sample.
	For the special case of a normal location model and a particular design of the subgroup, we show that the power improvement is equivalent to the power difference between a Monte Carlo $Z$-test and a Monte Carlo $t$-test.
\end{abstract}

	%
%

\textit{keywords: Group-invariance test; subgroup; permutation test; randomization test; resampling}
	
\end{titlepage}

	\section{Introduction}
	
	Permutation tests, randomization tests and related testing procedures are ubiquitous in modern-day statistical research \citep{onghena2018randomization, good2005permutation, berry2014chronicle}, for example in genomics \citep{tusher2001significance, li2013finding, debeer2020conditional}, neuroimaging \citep{eklund2016cluster} and economics \citep{young2019channeling}. 
	Such non-parametric (or semi-parametric) tests are useful in part because they require few assumptions on the data distribution \citep{anderson2001permutation, hemerik2021permutation}.
	Additionally, they have seen recent popularity in the simultaneous testing of many hypotheses, as they are often able to take into account the dependence structure of the data in an exact way, leading to relatively high power \citep{westfall1993resampling, tusher2001significance, meinshausen2006false, pesarin2010permutation, meinshausen2011asymptotic, hemerik2018false, blanchard2020post}. 
	For example, under strong positive dependence in the data, Bonferroni's multiple testing correction is very conservative and is greatly improved by a permutation method \citep{westfall1993resampling, westfall2008multiple}.
	
	These non-parametric tests often rely on the assumption that under the null hypothesis, the data distribution is invariant under a set $\calG$ of transformations that is a \emph{group}, in the algebraic sense \citep{lehmann2005testing, hemerik2018exact}.
	That is, every element in $\calG$ has an inverse in $\calG$ and $\calG$ is closed under composition.
	We will refer to tests based on a group-invariance assumption as \emph{group-invariance tests}.
	One prominent example is permutation tests.
	Another is sign-flipping tests \citep{fisher1935, efron1969student, bekker2008symmetry, davidson2008wild, winkler2014permutation, andreella2020permutation, blain2022notip, girardi2022post}.
	Sign-flipping is used, for instance, for testing in (generalized) linear models by sign-flipping residuals on score contributions \citep{hemerik2020robust, hemerik2021permutation, desantis2022inference}.
	But tests based other groups of transformations, e.g. rotations are also used \citep{langsrud2005rotation, solari2014rotation}.
	The requirement that $\calG$ is a group is fundamental; using a set of transformations that is not a group can lead to a very conservative or anti-conservative test \citep{southworth2009properties, hemerik2018exact, hemerik2021another}.

	\subsection{Current practice}	
		For moderate or large sample sizes, the cardinality $|\calG|$ is typically huge, so that it is often computationally infeasible to use the whole group.
		For example, the order of the permutation and sign-flipping groups is $n!$ and $2^n$, respectively, where $n$ is the number of observations.
		As a solution, it is universal practice among researchers to use a random finite subset of transformations \citep{eden1933validity, dwass1957, phipson2010permutation}.
		This can be done in such a way that the test is still exact \citep{hemerik2018exact}. 
		We will henceforth refer to such a test as a Monte Carlo group-invariance test.
		
		Using a small random subset transformations results in power loss compared to using the full group of transformations.
		Moreover, it leads to reduced replicability, since the test depends on the random subset of transformations that happens to be sampled. 
		For a Monte Carlo group-invariance test to have good power and to obtain replicable results, it is therefore important that a large random subset of transformations is used.
		Typically, this number is several times larger than $\alpha^{-1}$, where $\alpha$ is the nominal level of the test.
		For example, for a nominal level $\alpha = 0.05$, it is common to use 100-5000 random transformations. 
		Unfortunately, using a large random subset of transformations can remain prohibitive, as tests and multiple testing methods based on permutations (or other transformations) can be highly computationally intensive \citep{gao2010avoiding, kofler2012gowinda, hemerik2019permutation, vesely2021permutation}.
		
		To reduce the number of transformations required, a few methods have been proposed in the literature.
		For example, \citet{good2005permutation} approximates the permutation reference distribution using moment matching.
		\citet{winkler2016faster} review and propose additional methods for obtaining high-resolution \emph{p}-values based on a limited number of random permutations.
		However, although the resolution of permutation \emph{p}-values is improved, these approaches are not exact. 
		Moreover, the \emph{p}-values will still depend on the particular random sample of permutations that has been drawn, which also reduces the replicability of the results. 
		Further, it is not clear how to generally combine the methods in e.g. \citet{winkler2016faster} with permutation-based multiple testing methods, which are often not \emph{p}-value based.
		Thus, most of the drawbacks of the use of a limited set of random transformations have remained unresolved.

	\subsection{Contribution}
		In this paper, we propose an alternative approach to group-invariance testing: we replace the random subset of transformations with a (deterministic) \emph{subgroup} of transformations.
		We henceforth refer to such tests as \emph{subgroup-invariance tests}.
		Here, the subgroup can be selected in order to improve the power.
		If the subgroup is selected independently of the data, this does not affect the size of the test.
		This approach works without any additional assumptions and yields a fully replicable test.
		
		We illustrate this idea in a generalized location-shift model, which also contains the important two-sample comparison of means as a special case. 
		As the group $\calG$ we consider (subgroups of) the orthogonal group.
		This group contains the rotation group, the permutation group and the sign-flipping group as subgroups, and can be conveniently represented as a collection of orthonormal matrices.
		
		In this model, we compare the power of subgroup-invariance tests to the commonly used Monte Carlo group-invariance tests that are based on a random subset.
		We prove novel consistency results that link the consistency of a subgroup-invariance test to a real value $\delta_{\calS}$ that depends on the subgroup $\calS$ of $\calG$.
		Intuitively speaking, if $\delta_{\calS}$ is `large' then the elements of $\calS$ are more alike.
		We find that if $\delta_{\calS}$ is large, then a larger signal-to-noise ratio is required for the test to be consistent.
		Similarly, if $\delta_{\calS}$ is small, then a smaller signal-to-noise ratio is required for the test to be consistent.
		We prove an analogous result about the consistency of Monte Carlo group-invariance tests.
		There, the consistency turns out to depend on the full group $\calG$ through $\delta_{\calG}$.
		As we have $\delta_{\calS} \leq \delta_{\calG}$ for any subgroup $\calS$ of $\calG$, our results require a lower signal-to-noise ratio for consistency of subgroup-invariance tests than for the consistency of Monte Carlo group-invariance tests.
		
		Moreover, we provide a detailed power analysis in a normal location model, by linking the tests to $Z$-tests and $t$-tests.
		In particular, we consider subgroups $\calS$ for which $\delta_{\calS}^{\text{abs}} = 0$, a variant of $\delta_{\calS}$ we use for the analysis of two-sided tests.
		We show that subgroup-invariance tests based on such subgroups have the same size and power as a Monte Carlo $Z$-test.
		In addition, we show that a Monte Carlo orthogonal group-invariance test has the same size and power properties as a Monte Carlo $t$-test (regardless of the normality assumption).
		Under normality, Monte Carlo $Z$-tests are more powerful than Monte Carlo $t$-tests, if the same number of draws are used.
		As a consequence, the subgroup-invariance tests are more powerful, if the random sample and subgroup have the same number of elements.
		
		We also study the power through simulations, where we find that the subgroup-invariance tests we consider are substantially closer to the power of the test based on the full group, in comparison to the Monte Carlo tests based on the same number of transformations.
		In addition, to obtain the same power as a Monte Carlo test with $M$ random transformations, we require a subgroup with a cardinality smaller than $M$ (roughly $M/2$ in our simulation experiments).
		Thus, our method is an effective way to improve power or reduce the computation time; these are two sides of the same coin.	
		Moreover, simulations show that even if the subgroup is randomly chosen in some sense, the variability of the $p$-value (conditional on the data) is smaller than that of a $p$-value based on random sampling.

		Beside the properties of the tests, we also study the properties of the subgroups.
		We connect the problem of finding a subgroup for which $\delta_{\calS}$ is small to group code problems \citep{slepian1968group, conway1998sphere}, which relate to the spreading of points on a unit hypersphere.
		This connection exposes that demanding $\delta_{\calS}$ (or its two-sided equivalent $\delta_{\calS}^{\text{abs}}$) to be small, bounds the maximum order of the subgroup.
		For example, if $\delta_{\calS}^{\text{abs}} = 0$ then $|\calS| \leq n$, where $n$ is the dimension of the data.
		In addition, we provide an in-depth example of the subgroups of the sign-flipping group.

		Finally, we apply the methodology to fMRI data, to illustrate that our methodology can be combined with permutation-based (multiple) testing procedures.
		
		The appendix contains an example of the two-sample comparison problem, a description of an algorithm for constructing subgroups of the sign-flipping group, as well as (most of) the proofs of the results.
		The algorithm described in the Appendix is implemented by \citet{Koning_Near_Oracle_Subgroups_2022}, and \citet{Koning_Database_of_Near_2022} provides a readily available library of subgroups of the sign-flipping groups that were obtained using the algorithm.
		
	\section{Background: Testing invariance}
	In this section, we describe the problem of testing invariance and define group-invariance tests, to provide a context for our contributions.
	
	Let $\mathbb{R}^n$ be our sample space and $\calG$ be a compact group of $n \times n$ orthonormal matrices under matrix multiplication.
	For $\calG$ to be a group, it means that it contains the product of every pair of its elements, the inverse of every element, as well as the identity matrix $I$.
	If $\calG$ is finite, for example, then it is also compact.
	More generally, we can allow our sample space to be a topological space, and the group to be a compact topological group that acts continuously on the sample space (see e.g. \citealp{eaton1989group}).
		
	We observe a realization of a random variable $X$ in our sample space.
	The random variable $X$ is said to be $\calG$-invariant if $X \eqd GX$, for all $G \in \calG$.
	Equivalently, $X$ is $\calG$-invariant if $X \eqd \overline{G}X$, where $\overline{G}$ is uniform on $\calG$ independent of $X$ (see Appendix \ref{appn:invariance} for a proof).
	Our goal is to test whether $X$ is $\calG$-invariant at some level $\alpha \in (0, 1)$:
	\begin{align*}
		H_0 &: X \text{ is } \calG \text{-invariant}, \\
		H_1 &: X \text{ is not } \calG \text{-invariant}.
	\end{align*}
	
	\begin{exm}
		An important example is invariance under the permutation group: the collection of all permutation matrices $\calP$.
		Invariance under the permutation group sometimes referred to as `exchangeability'.
		As an example, if $X$ has i.i.d. elements then $X$ is exchangeable (though exchangeable vectors need not be i.i.d.).		
		Another example is the collection of `sign-flipping' matrices $\calR$, which are diagonal matrices with diagonal elements in $\{-1, 1\}$.
		Invariance under $\calR$ includes, for example, random vectors with independent elements and marginal distributions symmetric about 0.
		A further important example is the orthogonal group: the collection of all orthonormal matrices $\calH$, which includes all rotations.
		Random vectors that are $\calH$-invariant are also called `spherical'.
	\end{exm}
	
	\subsection{Group-invariance tests}
		In order to test invariance, it is standard practice to use a group-invariance test.
		Let $T : \calX \to \mathbb{R}$ act as a test statistic and $\overline{G}$ denote a random variable that is uniformly distributed on $\calG$.
		Then, a level $\alpha$ $\calG$-invariance test is defined as
		\begin{align*}
			\phi_{\alpha}^{\calG}(X) 
				&:= \mathbb{I}\{\SP_{\overline{G}}(T(\overline{G}X) \geq T(X)) \leq \alpha\},
		\end{align*}
		where $\SP_{\overline{G}}(T(\overline{G}X) \geq T(X))$ is the $p$-value of the test.
		It can be equivalently written in a form where $T(X)$ should exceed some threshold, as
		\begin{align*}
			\phi_{\alpha}^{\calG}(X)
				= \sI\{T(X) > q_{X}^{\alpha}(\calG)\},
		\end{align*}
		where $q_{X}^\alpha(\calG)$ is the $\alpha$-upper quantile of the distribution of $T(\overline{G}X)$, where $\overline{G}$ is uniform on $\calG$.
		Note here that the critical value $q_{X}^\alpha(\calG)$ depends on the data.
		
		Group-invariance tests are popular as they yield a test with exact size control (as opposed to approximate), regardless of the chosen test statistic.
		This is captured by the well-known
		\begin{thm}\label{thm:size_general}
			If $X$ is $\calG$-invariant, then $\SE_{X}\phi_{\alpha}^{\calG}(X) \leq \alpha$.
		\end{thm}
		For completeness, a proof can be found in Appendix \ref{appn:proofs}, together with all proofs of results that are not included in the text.
			
	\subsection{Practice}
		In practice, the groups under which invariance is tested are typically huge.
		This often makes it infeasible to compute the $p$-value $\SP_{\overline{G}}(T(\overline{G}X) \geq T(X))$.
		Therefore, it is universal practice to use a Monte Carlo (MC) test, instead.
		In particular, let $\calG_M$ be a set containing $M - 1$ independent draws from $\SP_{\overline{G}}$ and the identity $I$, and let $\overline{G}_M$ be uniform on $\calG_M$.
		Then, an $M$-draw MC $\calG$-invariance test is defined as
		\begin{align*}
			\phi_\alpha^{\calG_M}(X) 
				:= \mathbb{I}\{\SP_{\overline{G}_M}(T(\overline{G}_MX) \geq T(X)) \leq \alpha\}.
		\end{align*}
		
		This test is still exact \citep{hemerik2018exact}.
		However, it comes with two issues.
		First, if $M$ is small, then one would expect the power to be low.
		In addition, it is a random test as it depends on the random draw of $\calG_M$.
		This means that the outcome is not replicable, if one does not have access to the seed and random number generator.
		The replicability worsens if $M$ is small.
		
	\section{Subgroup-invariance tests}
	The key idea in this paper is the use of \emph{subgroup-invariance tests} as an alternative to Monte Carlo group-invariance tests.
	The construction of these tests relies on
		
	\begin{thm}\label{thm:subgroup}
		If $\calS$ is a subgroup of $\calG$ and $X$ is $\calG$-invariant, then $X$ is also $\calS$-invariant.
	\end{thm}
	
	The result follows immediately from the definition of $\calG$-invariance.
	If the subgroup $\calS$ is compact, we can use it to construct a subgroup-invariance test $\phi_{\alpha}^{\calS}$.
	Such a subgroup-invariance test controls size, by
	\begin{cor}\label{cor:subgroup-size}
		If $X$ is $\calG$-invariant, then $\SE_{X}\phi_{\alpha}^{\calS}(X) \leq \alpha$.
	\end{cor}
	\begin{proof}
		From Theorem \ref{thm:subgroup}, we learn that $X$ is $\calS$-invariant.
		Theorem \ref{thm:size_general} then yields $\SE_{X}\phi_{\alpha}^{\calS}(X) \leq \alpha$.	
	\end{proof}

	As as \emph{any} (compact) subgroup yields a test with exact size control, we can choose a subgroup that yields a test with desirable power properties.
	In particular, we will consider choosing a finite subgroup of order $M$, say, that yields a test with desirable power properties. 
	This way, we aim to obtain tests that are more powerful than MC group-invariance tests.
	
	In contrast to an MC group-invariance test, such a subgroup-invariance test has the additional benefit that it is completely deterministic given the data, and therefore does not suffer from replicability issues.
	If desirable, one can also construct Monte Carlo tests based on the subgroup, by sampling independently from a uniform distribution on the subgroup.

	While it is possible to beat MC group-invariance tests, as we shall see in the upcoming sections, it is impossible to find a subgroup-invariance test that is more powerful than the (full) group-invariance test by	
	\begin{thm}\label{thm:subgroup_vs_full_group}
		We have $\phi_{\alpha}^{\calG} \geq \phi_{\alpha}^{\calS}$.
	\end{thm}

	\section{Generalized location model}\label{sec:location-model}
	From this section onwards, we study subgroup-invariance tests in a generalized location model.
	Suppose that $X$ can be decomposed as
	\begin{align*}
		X = \iota\mu + \epsi,
	\end{align*}
	where $\iota$ is a unit $n$-vector, $\mu \in \mathbb{R}$ is the parameter of interest and $\epsi$ is a $\calG$-invariant random $n$-vector.
	As $\epsi$ is $\calG$-invariant, we have $\epsi \eqd \overline{G}\epsi$ where $\overline{G}$ is uniform on $\calG$, independent from $\epsi$.
	Hence, we can equivalently define $X$ as
	\begin{align*}
		X = \iota\mu + \overline{G}\epsi.
	\end{align*}	
	We sometimes write $X_{\overline{G}}$ to emphasize the dependence of $X$ on $\overline{G}$.
	
	We are interested in testing $H_0: \mu = 0$ at level $\alpha$.
	Notice that if $\mu = 0$ then $X = \epsi$, so that $X$ is $\calG$-invariant under $H_0$.
	As a consequence, if we test for $\calG$-invariance of $X$ and reject, we can also reject $H_0$.
	We consider both one-sided and two-sided alternatives $H_1 : \mu \neq 0$ and $H_1^+ : \mu > 0$, for which we use the test statistics $X \mapsto |\iota'X|$ and $X \mapsto \iota'X$, respectively.
	
	This model is more general than it may seem at first glance.
	In particular, choosing $\iota = n^{-1/2}(1, 1, \dots, 1)'$ yields a `standard' location model.
	But a different choice of $\iota$ can, for example, be used to model two-group comparisons of means: see Appendix \ref{appn:two_group}.
	In addition, the applicability of our results is not limited to the generalized location model.
	Indeed, many complex testing problems can be approximated by testing a location, such as inference in generalized linear models with nuisance covariates.
	There, a recent approach is to first compute residuals or score contributions and apply sign-flipping to those \citep{desantis2022inference, hemerik2020robust, hemerik2021permutation}.
	Such approaches are asymptotically exact.
	Thus, $\iota\mu + \epsi$ does not necessarily need to represent the `full' data distribution. 
	It can also represent a vector of test statistics as in \citet{andreella2020permutation}, or a vector of residuals as in e.g. \citet{winkler2014permutation}, or a vector of score contributions as in e.g. \citet{hemerik2020robust}.
	
	\subsection{The leak}
		The one-sided group-invariance test compares $\iota'X$ to the $\alpha$ upper-quantile of the distribution of
		\begin{align*}
			\iota'\overline{G}X = \iota'\overline{G}\iota\mu + \iota'\overline{G}\epsi,
		\end{align*}
		for fixed $X$.
		Note that this distribution depends on $\mu$ under the alternative, through the term $\iota'\overline{G}\iota\mu$.
		This term will play an important role in the remainder, and we follow an earlier version of \citet{dobriban2021consistency} in referring to it as a ``leak'' of signal into noise.
		
		To study the impact of the leak, we consider the following quantifications of its magnitude, for a given subset $\calS \neq \{I\}$ of the group of all orthonormal matrices $\calH$:
		\begin{align*}
			\delta_{\calS}
				&:= \sup_{S \in \calS\setminus\{I\}} \iota'S\iota,
		\end{align*}
		As an example, if each $S \in \calS \setminus\{I\}$ is a diagonal matrix with $n/2$ diagonal entries equal to $1$ and $n/2$ diagonal entries equal to 1, and $\iota = n^{-1/2}(1, 1, \dots, 1)'$, then $\delta_{\calS} = 0$.
		Subgroups of this form are studied in Section \ref{sec:sign-flipping}.	
		
	\subsection{Consistency and the leak}
		In this section, we describe conditions for non-asymptotic consistency of subgroup-invariance tests.		
		In particular, Theorem \ref{thm:root2-tech} shows that the magnitude of the leak has a negative impact on the power of the test.
		Specifically, a test based on a subgroup $\calS$ with a larger leak $\delta_{\calS}$ requires a larger signal-to-noise ratio $\mu / \|\epsi\|_2$ to be consistent.
		As a consequence, we would like to select a subgroup $\calS$ for which $\delta_{\calS}$ is minimized.
		
		\begin{thm}\label{thm:root2-tech}
			Let $\epsi$ be $\calG$-invariant and let $\calS$ be a finite subset of $\calG$.
			
			\begin{itemize}
				\item If $\alpha \geq 1 / |\calS|$ and $\mu\sqrt{1 - \delta_{\calS}} > \sqrt{2}\|\epsi\|_2$, then $\SE_{\overline{G}}\phi_{\alpha}^{\calS}(X_{\overline{G}} ) = 1$.
				\item If $\alpha = 1 / |\calS|$, $\epsi$ is $\calH$-invariant and $\delta_{\calS} < 1$, then $\mu\sqrt{1 - \delta_{\calS}} \geq \sqrt{2}\|\epsi\|_2$ if and only if $\SE_{\overline{H}} \phi_{\alpha}^{\calS}(X_{\overline{H}} ) = 1$.
			\end{itemize}
		\end{thm}

		Notice that the second claim shows that the result is `sharp', in the sense that there exists an $\alpha \geq 1 / |\calS|$ and a group $\calG$ such that there is equivalence.
		In Remark \ref{rmk:sharpness} in Appendix \ref{appn:proofs} we further discuss the sharpness of the first claim.
		There, we also provide a `sharper' claim that depends on $\calG$, but which is unfortunately harder to interpret as it no longer explicitly depends on the $\delta_{\calS}$.
		Furthermore, in Remark \ref{rmk:root2-discussion} in Appendix \ref{appn:proofs}, we also include a technical discussion of the claims of Theorem \ref{thm:root2-tech}.
		
		Interestingly, Theorem \ref{thm:root2-tech} does not require $\calS$ to be a sub\emph{group} of $\calG$.
		However, if $\calS$ is not a subgroup of $\calG$, we would generally not expect $\phi_{\alpha}^{\calS}$ to control size.

		A result about asymptotic consistency is easily obtained from Theorem \ref{thm:root2-tech}.
		\begin{cor}
			Let $X_{\overline{G}_m}^m$, $\iota_m$, $\mu_m$, $\epsi_m$, $\calS_m$, $\alpha_m$, $\calG_m$, $\overline{G}_m$ be sequences of the corresponding objects in Theorem \ref{thm:root2-tech} indexed by $m \in \mathbb{N}$.	
			Suppose $\alpha_m \geq 1 / |\calS_m|$ and $\tfrac{\mu_m}{\|\epsi_m\|_2}\sqrt{1 - \delta_{\calS_m}} \to c \in (\sqrt{2}, \infty]$, as $m \to \infty$.
			Then, $\SE_{\overline{G}_m}\phi_{\alpha_m}^{\calS_m}(X_{\overline{G}_m}^m) \to 1$.
		\end{cor}

		For the two-sided test, an analogous result holds.
		Here, we can define a two-sided version of the leak $\delta_{\calS}^{\text{abs}} = \sup_{S \in \calS\setminus\{I\}} |\iota'S\iota|$.		
		Denoting the two-sided analogue of $\phi_{\alpha}^{\calS}$ by $\overline{\phi}_{\alpha}^{\calS}$, we obtain
		\begin{thm}\label{thm:root2-tech-abs}
			Let $\epsi$ be $\calG$-invariant and $\calS$ be a finite subset of $\calG$.
			\begin{itemize}
				\item If $\alpha \geq 1 / |\calS|$ and $|\mu|\sqrt{1 - \delta_{\calS}^{\text{abs}}} > \sqrt{2}\|\epsi\|_2$, then $\SE_{\overline{G}} \overline{\phi}_{\alpha}^{\calS}(X_{\overline{G}}) = 1$.
				\item If $\alpha = 1 / |\calS|$, $\epsi$ is $\calH$-invariant and $\delta_{\calS}^{\text{abs}} < 1$, then $|\mu|\sqrt{1 - \delta_{\calS}^{\text{abs}}} \geq \sqrt{2}\|\epsi\|_2$ if and only if $\SE_{\overline{H}} \overline{\phi}_{\alpha}^{\calS}(X_{\overline{H}}) = 1$.
			\end{itemize}
		\end{thm}	
		
\section{Generalized location model: comparing the power of subgroup and Monte Carlo tests}
	In this section, we study and compare the power properties of the subgroup-invariance and MC group-invariance tests.
	We first compare the (non-asymptotic) consistency of the tests, through a consistency result for MC group-invariance tests that is analogous to Theorem \ref{thm:root2-tech}.
	Subsequently, we provide a more detailed power comparison under normality.
			
	\subsection{Comparison based on consistency}\label{sec:general_power}	
		To compare Monte Carlo group-invariance tests and subgroup-invariance tests, we derive an analogue of the consistency result in Theorem \ref{thm:root2-tech} for Monte Carlo tests.
		A similar analogue can be constructed for the two-sided Theorem \ref{thm:root2-tech-abs}.
		
		Theorem \ref{thm:root2-MC} shows that the magnitude of the leak $\delta_{\calG}$ of the group $\calG$ from which the Monte Carlo sample $\calG_M$ is taken, determines whether the Monte Carlo test is consistent.
			
		\begin{thm}\label{thm:root2-MC}
			Let $\epsi$ be $\calG$-invariant, $\|\epsi\|_2 > 0$ .
			Let the set $\calG_M$ consist of $M - 1$ uniformly drawn random variables from $\calG$ without replacement, and the identity.
			\begin{itemize}
				\item If $\alpha \geq 1 / M$ and $\mu\sqrt{1 - \delta_{\calG}} > \sqrt{2}\|\epsi\|_2$, then $\SE_{\calG_M}\SE_{\overline{G}}\phi_{\alpha}^{\calG_M}(X_{\overline{G}}) = 1$.
				\item If $\alpha = 1 / M$ and $\epsi$ is $\calH$-invariant and $\delta_{\calG} < 1$, then $\mu\sqrt{1 - \delta_{\calG}} \geq \sqrt{2}\|\epsi\|_2$ if and only if $\SE_{\calG_M}\SE_{\overline{H}} \phi_{\alpha}^{\calG_M}(X_{\overline{H}}) = 1$.
			\end{itemize}
		\end{thm}
	
		For any subgroup $\calS$ of $\calG$, we have $\delta_{\calS} \leq \delta_{\calG}$.
		Comparing Theorem \ref{thm:root2-tech} and \ref{thm:root2-MC} then tells us that a subgroup-invariance test is consistent for a smaller signal-to-noise ratio than a Monte Carlo group-invariance test based on $|\calS|$ draws from $\calG$.
		A precise statement of this claim is captured in Proposition \ref{prp:MC_vs_subgroup}.
		As a consequence, we find that the subgroup based tests are more powerful than their Monte Carlo counterpart, as measured in terms of consistency.
		
		\begin{prp}\label{prp:MC_vs_subgroup}
			Suppose $\epsi$ is $\calH$-invariant, $\|\epsi\|_2 > 0$, and $\calS$ is a subgroup of $\calG$ with $|\calS| = M$.
			Then, $\phi_{1/M}^{\calS}$ is consistent for a strictly smaller value of the signal-to-noise ratio $\mu / \|\epsi\|_2$ than $\phi_{1/M}^{\calG_{M}}$ if and only if $\delta_{\calS} < \delta_{\calG}$.
			In addition, both tests control size.
		\end{prp}		
		
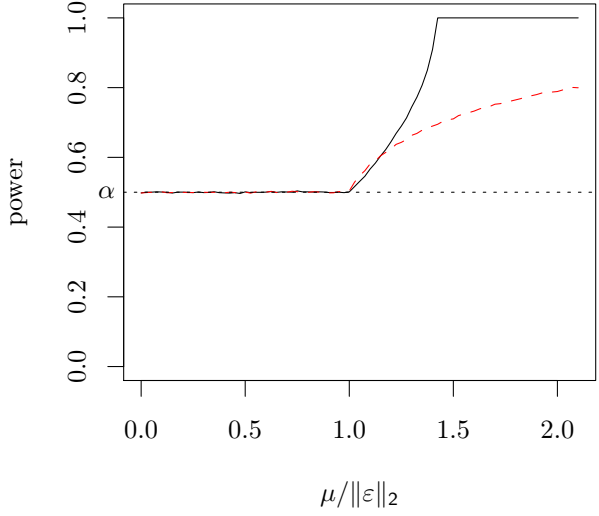
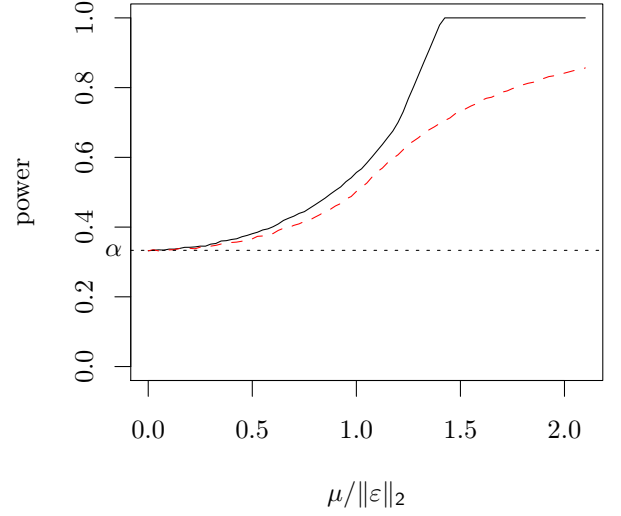
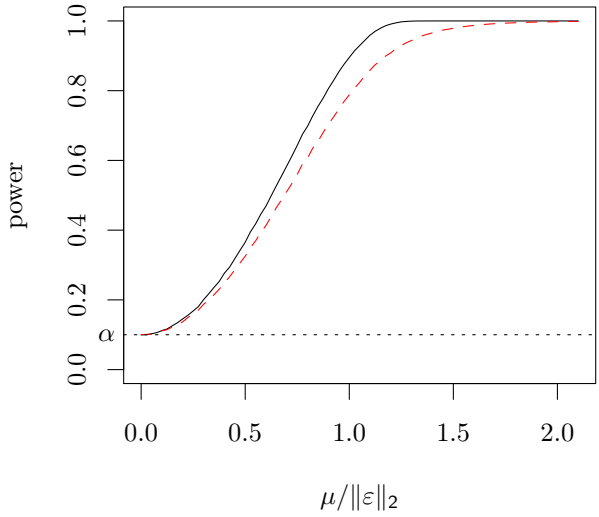
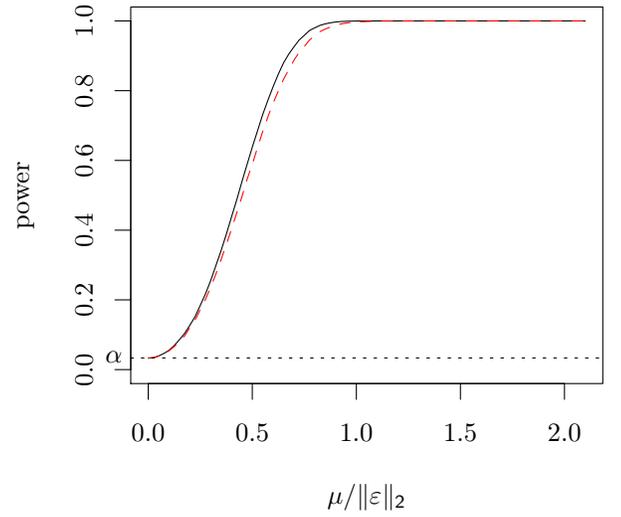
\begin{figure}[!htb]
			\begin{subfigure}[t]{0.4\textwidth}
		        \centering
\begin{tikzpicture}[x=1pt,y=1pt]
\definecolor{fillColor}{RGB}{255,255,255}
\path[use as bounding box,fill=fillColor,fill opacity=0.00] (0,0) rectangle (252.94,252.94);
\begin{scope}
\path[clip] (  0.00,  0.00) rectangle (252.94,252.94);
\definecolor{drawColor}{RGB}{0,0,0}

\path[draw=drawColor,line width= 0.4pt,line join=round,line cap=round] ( 55.81, 61.20) -- (213.26, 61.20);

\path[draw=drawColor,line width= 0.4pt,line join=round,line cap=round] ( 55.81, 61.20) -- ( 55.81, 55.20);

\path[draw=drawColor,line width= 0.4pt,line join=round,line cap=round] ( 95.17, 61.20) -- ( 95.17, 55.20);

\path[draw=drawColor,line width= 0.4pt,line join=round,line cap=round] (134.54, 61.20) -- (134.54, 55.20);

\path[draw=drawColor,line width= 0.4pt,line join=round,line cap=round] (173.90, 61.20) -- (173.90, 55.20);

\path[draw=drawColor,line width= 0.4pt,line join=round,line cap=round] (213.26, 61.20) -- (213.26, 55.20);

\node[text=drawColor,anchor=base,inner sep=0pt, outer sep=0pt, scale=  1.00] at ( 55.81, 39.60) {0.0};

\node[text=drawColor,anchor=base,inner sep=0pt, outer sep=0pt, scale=  1.00] at ( 95.17, 39.60) {0.5};

\node[text=drawColor,anchor=base,inner sep=0pt, outer sep=0pt, scale=  1.00] at (134.54, 39.60) {1.0};

\node[text=drawColor,anchor=base,inner sep=0pt, outer sep=0pt, scale=  1.00] at (173.90, 39.60) {1.5};

\node[text=drawColor,anchor=base,inner sep=0pt, outer sep=0pt, scale=  1.00] at (213.26, 39.60) {2.0};

\path[draw=drawColor,line width= 0.4pt,line join=round,line cap=round] ( 49.20, 66.48) -- ( 49.20,198.47);

\path[draw=drawColor,line width= 0.4pt,line join=round,line cap=round] ( 49.20, 66.48) -- ( 43.20, 66.48);

\path[draw=drawColor,line width= 0.4pt,line join=round,line cap=round] ( 49.20, 92.88) -- ( 43.20, 92.88);

\path[draw=drawColor,line width= 0.4pt,line join=round,line cap=round] ( 49.20,119.27) -- ( 43.20,119.27);

\path[draw=drawColor,line width= 0.4pt,line join=round,line cap=round] ( 49.20,145.67) -- ( 43.20,145.67);

\path[draw=drawColor,line width= 0.4pt,line join=round,line cap=round] ( 49.20,172.07) -- ( 43.20,172.07);

\path[draw=drawColor,line width= 0.4pt,line join=round,line cap=round] ( 49.20,198.47) -- ( 43.20,198.47);

\node[text=drawColor,rotate= 90.00,anchor=base,inner sep=0pt, outer sep=0pt, scale=  1.00] at ( 34.80, 66.48) {0.0};

\node[text=drawColor,rotate= 90.00,anchor=base,inner sep=0pt, outer sep=0pt, scale=  1.00] at ( 34.80, 92.88) {0.2};

\node[text=drawColor,rotate= 90.00,anchor=base,inner sep=0pt, outer sep=0pt, scale=  1.00] at ( 34.80,119.27) {0.4};

\node[text=drawColor,rotate= 90.00,anchor=base,inner sep=0pt, outer sep=0pt, scale=  1.00] at ( 34.80,145.67) {0.6};

\node[text=drawColor,rotate= 90.00,anchor=base,inner sep=0pt, outer sep=0pt, scale=  1.00] at ( 34.80,172.07) {0.8};

\node[text=drawColor,rotate= 90.00,anchor=base,inner sep=0pt, outer sep=0pt, scale=  1.00] at ( 34.80,198.47) {1.0};

\node[text=drawColor,anchor=base,inner sep=0pt, outer sep=0pt, scale=  1.00] at ( 42.80,130.47) {$\alpha$};

\path[draw=drawColor,line width= 0.4pt,line join=round,line cap=round] ( 49.20, 61.20) --
	(227.75, 61.20) --
	(227.75,203.75) --
	( 49.20,203.75) --
	( 49.20, 61.20);
\end{scope}
\begin{scope}
\path[clip] (  0.00,  0.00) rectangle (252.94,252.94);
\definecolor{drawColor}{RGB}{0,0,0}

\node[text=drawColor,anchor=base,inner sep=0pt, outer sep=0pt, scale=  1.00] at (138.47, 15.60) {$\mu / \|\epsi\|_2$};

\node[text=drawColor,rotate= 90.00,anchor=base,inner sep=0pt, outer sep=0pt, scale=  1.00] at ( 10.80,132.47) {power};
\end{scope}
\begin{scope}
\path[clip] ( 49.20, 61.20) rectangle (227.75,203.75);
\definecolor{drawColor}{RGB}{0,0,0}

\path[draw=drawColor,line width= 0.4pt,line join=round,line cap=round] ( 55.81,132.22) --
	( 57.78,132.48) --
	( 59.75,132.49) --
	( 61.72,132.63) --
	( 63.69,132.53) --
	( 65.65,132.43) --
	( 67.62,132.16) --
	( 69.59,132.51) --
	( 71.56,132.51) --
	( 73.53,132.44) --
	( 75.49,132.31) --
	( 77.46,132.55) --
	( 79.43,132.33) --
	( 81.40,132.38) --
	( 83.37,132.66) --
	( 85.33,132.30) --
	( 87.30,132.21) --
	( 89.27,132.14) --
	( 91.24,132.31) --
	( 93.21,132.00) --
	( 95.17,132.61) --
	( 97.14,132.33) --
	( 99.11,132.44) --
	(101.08,132.44) --
	(103.05,132.51) --
	(105.01,132.73) --
	(106.98,132.42) --
	(108.95,132.62) --
	(110.92,132.56) --
	(112.89,132.49) --
	(114.86,132.98) --
	(116.82,132.61) --
	(118.79,132.61) --
	(120.76,132.53) --
	(122.73,132.63) --
	(124.70,132.57) --
	(126.66,132.37) --
	(128.63,132.37) --
	(130.60,132.48) --
	(132.57,132.31) --
	(134.54,132.60) --
	(136.50,134.57) --
	(138.47,136.58) --
	(140.44,138.52) --
	(142.41,141.24) --
	(144.38,143.47) --
	(146.34,146.17) --
	(148.31,148.61) --
	(150.28,151.53) --
	(152.25,154.73) --
	(154.22,157.50) --
	(156.19,160.61) --
	(158.15,164.75) --
	(160.12,168.45) --
	(162.09,172.98) --
	(164.06,178.61) --
	(166.03,186.52) --
	(167.99,198.47) --
	(169.96,198.47) --
	(171.93,198.47) --
	(173.90,198.47) --
	(175.87,198.47) --
	(177.83,198.47) --
	(179.80,198.47) --
	(181.77,198.47) --
	(183.74,198.47) --
	(185.71,198.47) --
	(187.67,198.47) --
	(189.64,198.47) --
	(191.61,198.47) --
	(193.58,198.47) --
	(195.55,198.47) --
	(197.52,198.47) --
	(199.48,198.47) --
	(201.45,198.47) --
	(203.42,198.47) --
	(205.39,198.47) --
	(207.36,198.47) --
	(209.32,198.47) --
	(211.29,198.47) --
	(213.26,198.47) --
	(215.23,198.47) --
	(217.20,198.47) --
	(219.16,198.47) --
	(221.13,198.47);
\definecolor{drawColor}{RGB}{255,0,0}

\path[draw=drawColor,line width= 0.4pt,dash pattern=on 4pt off 4pt ,line join=round,line cap=round] ( 55.81,132.18) --
	( 57.78,132.34) --
	( 59.75,132.45) --
	( 61.72,132.31) --
	( 63.69,132.62) --
	( 65.65,132.26) --
	( 67.62,132.36) --
	( 69.59,132.62) --
	( 71.56,132.45) --
	( 73.53,132.34) --
	( 75.49,132.35) --
	( 77.46,132.55) --
	( 79.43,132.46) --
	( 81.40,132.41) --
	( 83.37,132.45) --
	( 85.33,132.34) --
	( 87.30,132.29) --
	( 89.27,132.46) --
	( 91.24,132.36) --
	( 93.21,131.72) --
	( 95.17,132.54) --
	( 97.14,132.14) --
	( 99.11,132.59) --
	(101.08,132.64) --
	(103.05,132.48) --
	(105.01,132.80) --
	(106.98,132.34) --
	(108.95,132.85) --
	(110.92,132.47) --
	(112.89,132.84) --
	(114.86,132.67) --
	(116.82,132.41) --
	(118.79,132.38) --
	(120.76,132.66) --
	(122.73,132.55) --
	(124.70,132.44) --
	(126.66,132.22) --
	(128.63,132.26) --
	(130.60,132.70) --
	(132.57,132.94) --
	(134.54,132.97) --
	(136.50,136.23) --
	(138.47,138.77) --
	(140.44,140.62) --
	(142.41,143.18) --
	(144.38,144.27) --
	(146.34,146.25) --
	(148.31,147.53) --
	(150.28,148.86) --
	(152.25,150.59) --
	(154.22,151.35) --
	(156.19,152.42) --
	(158.15,154.02) --
	(160.12,154.73) --
	(162.09,155.96) --
	(164.06,156.86) --
	(166.03,157.51) --
	(167.99,158.20) --
	(169.96,159.32) --
	(171.93,159.93) --
	(173.90,160.23) --
	(175.87,161.49) --
	(177.83,162.06) --
	(179.80,162.62) --
	(181.77,163.19) --
	(183.74,163.98) --
	(185.71,164.64) --
	(187.67,164.93) --
	(189.64,165.79) --
	(191.61,165.99) --
	(193.58,166.58) --
	(195.55,166.96) --
	(197.52,167.44) --
	(199.48,167.92) --
	(201.45,168.80) --
	(203.42,168.99) --
	(205.39,169.43) --
	(207.36,170.03) --
	(209.32,169.93) --
	(211.29,170.40) --
	(213.26,170.59) --
	(215.23,171.18) --
	(217.20,171.28) --
	(219.16,172.13) --
	(221.13,172.01);
\definecolor{drawColor}{RGB}{0,0,0}

\path[draw=drawColor,line width= 0.4pt,dash pattern=on 1pt off 3pt ,line join=round,line cap=round] ( 49.20,132.47) -- (227.75,132.47);
\end{scope}
\end{tikzpicture}
		        \caption{Power for $n = 2$, with $\alpha = 1/2$.} \label{fig:A}
		    \end{subfigure}
		    \hfill
		    \begin{subfigure}[t]{0.4\textwidth}
		        \centering
\begin{tikzpicture}[x=1pt,y=1pt]
\definecolor{fillColor}{RGB}{255,255,255}
\path[use as bounding box,fill=fillColor,fill opacity=0.00] (0,0) rectangle (252.94,252.94);
\begin{scope}
\path[clip] (  0.00,  0.00) rectangle (252.94,252.94);
\definecolor{drawColor}{RGB}{0,0,0}

\path[draw=drawColor,line width= 0.4pt,line join=round,line cap=round] ( 55.81, 61.20) -- (213.26, 61.20);

\path[draw=drawColor,line width= 0.4pt,line join=round,line cap=round] ( 55.81, 61.20) -- ( 55.81, 55.20);

\path[draw=drawColor,line width= 0.4pt,line join=round,line cap=round] ( 95.17, 61.20) -- ( 95.17, 55.20);

\path[draw=drawColor,line width= 0.4pt,line join=round,line cap=round] (134.54, 61.20) -- (134.54, 55.20);

\path[draw=drawColor,line width= 0.4pt,line join=round,line cap=round] (173.90, 61.20) -- (173.90, 55.20);

\path[draw=drawColor,line width= 0.4pt,line join=round,line cap=round] (213.26, 61.20) -- (213.26, 55.20);

\node[text=drawColor,anchor=base,inner sep=0pt, outer sep=0pt, scale=  1.00] at ( 55.81, 39.60) {0.0};

\node[text=drawColor,anchor=base,inner sep=0pt, outer sep=0pt, scale=  1.00] at ( 95.17, 39.60) {0.5};

\node[text=drawColor,anchor=base,inner sep=0pt, outer sep=0pt, scale=  1.00] at (134.54, 39.60) {1.0};

\node[text=drawColor,anchor=base,inner sep=0pt, outer sep=0pt, scale=  1.00] at (173.90, 39.60) {1.5};

\node[text=drawColor,anchor=base,inner sep=0pt, outer sep=0pt, scale=  1.00] at (213.26, 39.60) {2.0};

\path[draw=drawColor,line width= 0.4pt,line join=round,line cap=round] ( 49.20, 66.48) -- ( 49.20,198.47);

\path[draw=drawColor,line width= 0.4pt,line join=round,line cap=round] ( 49.20, 66.48) -- ( 43.20, 66.48);

\path[draw=drawColor,line width= 0.4pt,line join=round,line cap=round] ( 49.20, 92.88) -- ( 43.20, 92.88);

\path[draw=drawColor,line width= 0.4pt,line join=round,line cap=round] ( 49.20,119.27) -- ( 43.20,119.27);

\path[draw=drawColor,line width= 0.4pt,line join=round,line cap=round] ( 49.20,145.67) -- ( 43.20,145.67);

\path[draw=drawColor,line width= 0.4pt,line join=round,line cap=round] ( 49.20,172.07) -- ( 43.20,172.07);

\path[draw=drawColor,line width= 0.4pt,line join=round,line cap=round] ( 49.20,198.47) -- ( 43.20,198.47);

\node[text=drawColor,rotate= 90.00,anchor=base,inner sep=0pt, outer sep=0pt, scale=  1.00] at ( 34.80, 66.48) {0.0};

\node[text=drawColor,rotate= 90.00,anchor=base,inner sep=0pt, outer sep=0pt, scale=  1.00] at ( 34.80, 92.88) {0.2};

\node[text=drawColor,rotate= 90.00,anchor=base,inner sep=0pt, outer sep=0pt, scale=  1.00] at ( 34.80,119.27) {0.4};

\node[text=drawColor,rotate= 90.00,anchor=base,inner sep=0pt, outer sep=0pt, scale=  1.00] at ( 34.80,145.67) {0.6};

\node[text=drawColor,rotate= 90.00,anchor=base,inner sep=0pt, outer sep=0pt, scale=  1.00] at ( 34.80,172.07) {0.8};

\node[text=drawColor,rotate= 90.00,anchor=base,inner sep=0pt, outer sep=0pt, scale=  1.00] at ( 34.80,198.47) {1.0};

\node[text=drawColor,anchor=base,inner sep=0pt, outer sep=0pt, scale=  1.00] at ( 42.80,108.47) {$\alpha$};

\path[draw=drawColor,line width= 0.4pt,line join=round,line cap=round] ( 49.20, 61.20) --
	(227.75, 61.20) --
	(227.75,203.75) --
	( 49.20,203.75) --
	( 49.20, 61.20);
\end{scope}
\begin{scope}
\path[clip] (  0.00,  0.00) rectangle (252.94,252.94);
\definecolor{drawColor}{RGB}{0,0,0}

\node[text=drawColor,anchor=base,inner sep=0pt, outer sep=0pt, scale=  1.00] at (138.47, 15.60) {$\mu/\|\epsi\|_2$};

\node[text=drawColor,rotate= 90.00,anchor=base,inner sep=0pt, outer sep=0pt, scale=  1.00] at ( 10.80,132.47) {power};
\end{scope}
\begin{scope}
\path[clip] ( 49.20, 61.20) rectangle (227.75,203.75);
\definecolor{drawColor}{RGB}{0,0,0}

\path[draw=drawColor,line width= 0.4pt,line join=round,line cap=round] ( 55.81,110.32) --
	( 57.78,110.49) --
	( 59.75,110.66) --
	( 61.72,110.56) --
	( 63.69,110.94) --
	( 65.65,110.93) --
	( 67.62,111.18) --
	( 69.59,111.61) --
	( 71.56,111.63) --
	( 73.53,111.77) --
	( 75.49,112.07) --
	( 77.46,112.05) --
	( 79.43,112.82) --
	( 81.40,113.07) --
	( 83.37,114.01) --
	( 85.33,114.13) --
	( 87.30,114.58) --
	( 89.27,114.85) --
	( 91.24,115.62) --
	( 93.21,116.12) --
	( 95.17,116.77) --
	( 97.14,117.33) --
	( 99.11,118.19) --
	(101.08,118.60) --
	(103.05,119.42) --
	(105.01,120.46) --
	(106.98,121.80) --
	(108.95,122.65) --
	(110.92,123.42) --
	(112.89,124.46) --
	(114.86,125.11) --
	(116.82,126.38) --
	(118.79,127.64) --
	(120.76,128.94) --
	(122.73,130.25) --
	(124.70,131.77) --
	(126.66,133.17) --
	(128.63,134.50) --
	(130.60,136.46) --
	(132.57,137.92) --
	(134.54,139.94) --
	(136.50,141.34) --
	(138.47,143.52) --
	(140.44,145.81) --
	(142.41,148.19) --
	(144.38,150.55) --
	(146.34,153.10) --
	(148.31,155.59) --
	(150.28,158.87) --
	(152.25,162.93) --
	(154.22,167.94) --
	(156.19,172.42) --
	(158.15,177.11) --
	(160.12,181.80) --
	(162.09,186.37) --
	(164.06,191.06) --
	(166.03,195.78) --
	(167.99,198.47) --
	(169.96,198.47) --
	(171.93,198.47) --
	(173.90,198.47) --
	(175.87,198.47) --
	(177.83,198.47) --
	(179.80,198.47) --
	(181.77,198.47) --
	(183.74,198.47) --
	(185.71,198.47) --
	(187.67,198.47) --
	(189.64,198.47) --
	(191.61,198.47) --
	(193.58,198.47) --
	(195.55,198.47) --
	(197.52,198.47) --
	(199.48,198.47) --
	(201.45,198.47) --
	(203.42,198.47) --
	(205.39,198.47) --
	(207.36,198.47) --
	(209.32,198.47) --
	(211.29,198.47) --
	(213.26,198.47) --
	(215.23,198.47) --
	(217.20,198.47) --
	(219.16,198.47) --
	(221.13,198.47);
\definecolor{drawColor}{RGB}{255,0,0}

\path[draw=drawColor,line width= 0.4pt,dash pattern=on 4pt off 4pt ,line join=round,line cap=round] ( 55.81,110.31) --
	( 57.78,110.69) --
	( 59.75,110.26) --
	( 61.72,110.68) --
	( 63.69,110.64) --
	( 65.65,110.73) --
	( 67.62,111.16) --
	( 69.59,111.09) --
	( 71.56,111.23) --
	( 73.53,111.07) --
	( 75.49,111.68) --
	( 77.46,111.93) --
	( 79.43,112.06) --
	( 81.40,112.36) --
	( 83.37,112.87) --
	( 85.33,113.06) --
	( 87.30,113.45) --
	( 89.27,113.59) --
	( 91.24,113.92) --
	( 93.21,114.65) --
	( 95.17,114.82) --
	( 97.14,115.77) --
	( 99.11,115.91) --
	(101.08,116.13) --
	(103.05,116.84) --
	(105.01,118.02) --
	(106.98,118.83) --
	(108.95,119.11) --
	(110.92,119.94) --
	(112.89,120.50) --
	(114.86,121.42) --
	(116.82,121.91) --
	(118.79,122.98) --
	(120.76,123.99) --
	(122.73,125.20) --
	(124.70,126.01) --
	(126.66,127.29) --
	(128.63,128.20) --
	(130.60,129.91) --
	(132.57,130.84) --
	(134.54,132.78) --
	(136.50,134.35) --
	(138.47,135.91) --
	(140.44,138.18) --
	(142.41,140.22) --
	(144.38,141.61) --
	(146.34,143.64) --
	(148.31,145.29) --
	(150.28,146.62) --
	(152.25,148.57) --
	(154.22,150.58) --
	(156.19,151.85) --
	(158.15,153.19) --
	(160.12,154.97) --
	(162.09,156.13) --
	(164.06,157.23) --
	(166.03,158.50) --
	(167.99,159.60) --
	(169.96,160.60) --
	(171.93,162.18) --
	(173.90,163.04) --
	(175.87,164.05) --
	(177.83,165.08) --
	(179.80,165.93) --
	(181.77,167.19) --
	(183.74,168.04) --
	(185.71,168.44) --
	(187.67,169.39) --
	(189.64,170.34) --
	(191.61,170.83) --
	(193.58,171.72) --
	(195.55,172.30) --
	(197.52,173.12) --
	(199.48,173.71) --
	(201.45,174.10) --
	(203.42,174.57) --
	(205.39,175.54) --
	(207.36,176.25) --
	(209.32,176.55) --
	(211.29,177.19) --
	(213.26,177.56) --
	(215.23,178.09) --
	(217.20,178.63) --
	(219.16,178.93) --
	(221.13,179.52);
\definecolor{drawColor}{RGB}{0,0,0}

\path[draw=drawColor,line width= 0.4pt,dash pattern=on 1pt off 3pt ,line join=round,line cap=round] ( 49.20,110.47) -- (227.75,110.47);
\end{scope}
\end{tikzpicture}
		        \caption{Power for $n = 3$, with $\alpha = 1/3$.} \label{fig:B}
		    \end{subfigure} \\
		    
		    \begin{subfigure}[t]{0.4\textwidth}
		        \centering
\begin{tikzpicture}[x=1pt,y=1pt]
\definecolor{fillColor}{RGB}{255,255,255}
\path[use as bounding box,fill=fillColor,fill opacity=0.00] (0,0) rectangle (252.94,252.94);
\begin{scope}
\path[clip] (  0.00,  0.00) rectangle (252.94,252.94);
\definecolor{drawColor}{RGB}{0,0,0}

\path[draw=drawColor,line width= 0.4pt,line join=round,line cap=round] ( 55.81, 61.20) -- (213.26, 61.20);

\path[draw=drawColor,line width= 0.4pt,line join=round,line cap=round] ( 55.81, 61.20) -- ( 55.81, 55.20);

\path[draw=drawColor,line width= 0.4pt,line join=round,line cap=round] ( 95.17, 61.20) -- ( 95.17, 55.20);

\path[draw=drawColor,line width= 0.4pt,line join=round,line cap=round] (134.54, 61.20) -- (134.54, 55.20);

\path[draw=drawColor,line width= 0.4pt,line join=round,line cap=round] (173.90, 61.20) -- (173.90, 55.20);

\path[draw=drawColor,line width= 0.4pt,line join=round,line cap=round] (213.26, 61.20) -- (213.26, 55.20);

\node[text=drawColor,anchor=base,inner sep=0pt, outer sep=0pt, scale=  1.00] at ( 55.81, 39.60) {0.0};

\node[text=drawColor,anchor=base,inner sep=0pt, outer sep=0pt, scale=  1.00] at ( 95.17, 39.60) {0.5};

\node[text=drawColor,anchor=base,inner sep=0pt, outer sep=0pt, scale=  1.00] at (134.54, 39.60) {1.0};

\node[text=drawColor,anchor=base,inner sep=0pt, outer sep=0pt, scale=  1.00] at (173.90, 39.60) {1.5};

\node[text=drawColor,anchor=base,inner sep=0pt, outer sep=0pt, scale=  1.00] at (213.26, 39.60) {2.0};

\path[draw=drawColor,line width= 0.4pt,line join=round,line cap=round] ( 49.20, 66.48) -- ( 49.20,198.47);

\path[draw=drawColor,line width= 0.4pt,line join=round,line cap=round] ( 49.20, 66.48) -- ( 43.20, 66.48);

\path[draw=drawColor,line width= 0.4pt,line join=round,line cap=round] ( 49.20, 92.88) -- ( 43.20, 92.88);

\path[draw=drawColor,line width= 0.4pt,line join=round,line cap=round] ( 49.20,119.27) -- ( 43.20,119.27);

\path[draw=drawColor,line width= 0.4pt,line join=round,line cap=round] ( 49.20,145.67) -- ( 43.20,145.67);

\path[draw=drawColor,line width= 0.4pt,line join=round,line cap=round] ( 49.20,172.07) -- ( 43.20,172.07);

\path[draw=drawColor,line width= 0.4pt,line join=round,line cap=round] ( 49.20,198.47) -- ( 43.20,198.47);

\node[text=drawColor,rotate= 90.00,anchor=base,inner sep=0pt, outer sep=0pt, scale=  1.00] at ( 34.80, 66.48) {0.0};

\node[text=drawColor,rotate= 90.00,anchor=base,inner sep=0pt, outer sep=0pt, scale=  1.00] at ( 34.80, 92.88) {0.2};

\node[text=drawColor,rotate= 90.00,anchor=base,inner sep=0pt, outer sep=0pt, scale=  1.00] at ( 34.80,119.27) {0.4};

\node[text=drawColor,rotate= 90.00,anchor=base,inner sep=0pt, outer sep=0pt, scale=  1.00] at ( 34.80,145.67) {0.6};

\node[text=drawColor,rotate= 90.00,anchor=base,inner sep=0pt, outer sep=0pt, scale=  1.00] at ( 34.80,172.07) {0.8};

\node[text=drawColor,rotate= 90.00,anchor=base,inner sep=0pt, outer sep=0pt, scale=  1.00] at ( 34.80,198.47) {1.0};

\node[text=drawColor,anchor=base,inner sep=0pt, outer sep=0pt, scale=  1.00] at ( 42.80,77.47) {$\alpha$};

\path[draw=drawColor,line width= 0.4pt,line join=round,line cap=round] ( 49.20, 61.20) --
	(227.75, 61.20) --
	(227.75,203.75) --
	( 49.20,203.75) --
	( 49.20, 61.20);
\end{scope}
\begin{scope}
\path[clip] (  0.00,  0.00) rectangle (252.94,252.94);
\definecolor{drawColor}{RGB}{0,0,0}

\node[text=drawColor,anchor=base,inner sep=0pt, outer sep=0pt, scale=  1.00] at (138.47, 15.60) {$\mu/\|\epsi\|_2$};

\node[text=drawColor,rotate= 90.00,anchor=base,inner sep=0pt, outer sep=0pt, scale=  1.00] at ( 10.80,132.47) {power};
\end{scope}
\begin{scope}
\path[clip] ( 49.20, 61.20) rectangle (227.75,203.75);
\definecolor{drawColor}{RGB}{0,0,0}

\path[draw=drawColor,line width= 0.4pt,line join=round,line cap=round] ( 55.81, 79.62) --
	( 57.78, 79.68) --
	( 59.75, 80.05) --
	( 61.72, 80.48) --
	( 63.69, 81.28) --
	( 65.65, 81.88) --
	( 67.62, 83.10) --
	( 69.59, 84.26) --
	( 71.56, 85.70) --
	( 73.53, 87.08) --
	( 75.49, 88.69) --
	( 77.46, 90.35) --
	( 79.43, 92.96) --
	( 81.40, 95.22) --
	( 83.37, 97.62) --
	( 85.33, 99.82) --
	( 87.30,102.99) --
	( 89.27,105.28) --
	( 91.24,108.48) --
	( 93.21,111.64) --
	( 95.17,114.70) --
	( 97.14,118.60) --
	( 99.11,121.64) --
	(101.08,125.33) --
	(103.05,128.42) --
	(105.01,132.31) --
	(106.98,136.13) --
	(108.95,139.86) --
	(110.92,143.63) --
	(112.89,147.42) --
	(114.86,151.37) --
	(116.82,155.56) --
	(118.79,158.66) --
	(120.76,162.62) --
	(122.73,166.18) --
	(124.70,169.32) --
	(126.66,172.89) --
	(128.63,175.96) --
	(130.60,179.23) --
	(132.57,182.08) --
	(134.54,184.63) --
	(136.50,187.15) --
	(138.47,189.16) --
	(140.44,191.16) --
	(142.41,193.10) --
	(144.38,194.56) --
	(146.34,195.72) --
	(148.31,196.67) --
	(150.28,197.30) --
	(152.25,197.75) --
	(154.22,198.12) --
	(156.19,198.29) --
	(158.15,198.40) --
	(160.12,198.45) --
	(162.09,198.46) --
	(164.06,198.47) --
	(166.03,198.47) --
	(167.99,198.47) --
	(169.96,198.47) --
	(171.93,198.47) --
	(173.90,198.47) --
	(175.87,198.47) --
	(177.83,198.47) --
	(179.80,198.47) --
	(181.77,198.47) --
	(183.74,198.47) --
	(185.71,198.47) --
	(187.67,198.47) --
	(189.64,198.47) --
	(191.61,198.47) --
	(193.58,198.47) --
	(195.55,198.47) --
	(197.52,198.47) --
	(199.48,198.47) --
	(201.45,198.47) --
	(203.42,198.47) --
	(205.39,198.47) --
	(207.36,198.47) --
	(209.32,198.47) --
	(211.29,198.47) --
	(213.26,198.47) --
	(215.23,198.47) --
	(217.20,198.47) --
	(219.16,198.47) --
	(221.13,198.47);
\definecolor{drawColor}{RGB}{255,0,0}

\path[draw=drawColor,line width= 0.4pt,dash pattern=on 4pt off 4pt ,line join=round,line cap=round] ( 55.81, 79.71) --
	( 57.78, 79.83) --
	( 59.75, 79.89) --
	( 61.72, 80.42) --
	( 63.69, 81.15) --
	( 65.65, 81.61) --
	( 67.62, 82.61) --
	( 69.59, 83.37) --
	( 71.56, 84.60) --
	( 73.53, 86.21) --
	( 75.49, 87.57) --
	( 77.46, 89.02) --
	( 79.43, 91.21) --
	( 81.40, 92.85) --
	( 83.37, 94.83) --
	( 85.33, 96.90) --
	( 87.30, 99.28) --
	( 89.27,101.64) --
	( 91.24,104.26) --
	( 93.21,106.86) --
	( 95.17,109.38) --
	( 97.14,112.66) --
	( 99.11,115.22) --
	(101.08,118.44) --
	(103.05,121.06) --
	(105.01,124.45) --
	(106.98,127.86) --
	(108.95,130.80) --
	(110.92,133.93) --
	(112.89,137.02) --
	(114.86,140.39) --
	(116.82,144.28) --
	(118.79,146.65) --
	(120.76,150.38) --
	(122.73,152.97) --
	(124.70,156.30) --
	(126.66,159.41) --
	(128.63,162.16) --
	(130.60,165.11) --
	(132.57,167.99) --
	(134.54,170.50) --
	(136.50,173.01) --
	(138.47,175.20) --
	(140.44,177.73) --
	(142.41,179.85) --
	(144.38,181.95) --
	(146.34,183.27) --
	(148.31,185.20) --
	(150.28,186.41) --
	(152.25,188.06) --
	(154.22,189.19) --
	(156.19,190.25) --
	(158.15,191.10) --
	(160.12,191.90) --
	(162.09,192.84) --
	(164.06,193.44) --
	(166.03,193.96) --
	(167.99,194.59) --
	(169.96,195.02) --
	(171.93,195.34) --
	(173.90,195.71) --
	(175.87,196.09) --
	(177.83,196.29) --
	(179.80,196.59) --
	(181.77,196.72) --
	(183.74,197.01) --
	(185.71,197.15) --
	(187.67,197.32) --
	(189.64,197.43) --
	(191.61,197.57) --
	(193.58,197.63) --
	(195.55,197.72) --
	(197.52,197.79) --
	(199.48,197.91) --
	(201.45,197.93) --
	(203.42,198.01) --
	(205.39,198.04) --
	(207.36,198.08) --
	(209.32,198.11) --
	(211.29,198.14) --
	(213.26,198.18) --
	(215.23,198.23) --
	(217.20,198.24) --
	(219.16,198.27) --
	(221.13,198.25);
\definecolor{drawColor}{RGB}{0,0,0}

\path[draw=drawColor,line width= 0.4pt,dash pattern=on 1pt off 3pt ,line join=round,line cap=round] ( 49.20, 79.68) -- (227.75, 79.68);
\end{scope}
\end{tikzpicture}
		        \caption{Power for $n = 10$, with $\alpha = 1/10$.} \label{fig:C}
		    \end{subfigure}
		    \hfill
		    \begin{subfigure}[t]{0.4\textwidth}
		        \centering
\begin{tikzpicture}[x=1pt,y=1pt]
\definecolor{fillColor}{RGB}{255,255,255}
\path[use as bounding box,fill=fillColor,fill opacity=0.00] (0,0) rectangle (252.94,252.94);
\begin{scope}
\path[clip] (  0.00,  0.00) rectangle (252.94,252.94);
\definecolor{drawColor}{RGB}{0,0,0}

\path[draw=drawColor,line width= 0.4pt,line join=round,line cap=round] ( 55.81, 61.20) -- (213.26, 61.20);

\path[draw=drawColor,line width= 0.4pt,line join=round,line cap=round] ( 55.81, 61.20) -- ( 55.81, 55.20);

\path[draw=drawColor,line width= 0.4pt,line join=round,line cap=round] ( 95.17, 61.20) -- ( 95.17, 55.20);

\path[draw=drawColor,line width= 0.4pt,line join=round,line cap=round] (134.54, 61.20) -- (134.54, 55.20);

\path[draw=drawColor,line width= 0.4pt,line join=round,line cap=round] (173.90, 61.20) -- (173.90, 55.20);

\path[draw=drawColor,line width= 0.4pt,line join=round,line cap=round] (213.26, 61.20) -- (213.26, 55.20);

\node[text=drawColor,anchor=base,inner sep=0pt, outer sep=0pt, scale=  1.00] at ( 55.81, 39.60) {0.0};

\node[text=drawColor,anchor=base,inner sep=0pt, outer sep=0pt, scale=  1.00] at ( 95.17, 39.60) {0.5};

\node[text=drawColor,anchor=base,inner sep=0pt, outer sep=0pt, scale=  1.00] at (134.54, 39.60) {1.0};

\node[text=drawColor,anchor=base,inner sep=0pt, outer sep=0pt, scale=  1.00] at (173.90, 39.60) {1.5};

\node[text=drawColor,anchor=base,inner sep=0pt, outer sep=0pt, scale=  1.00] at (213.26, 39.60) {2.0};

\path[draw=drawColor,line width= 0.4pt,line join=round,line cap=round] ( 49.20, 66.48) -- ( 49.20,198.47);

\path[draw=drawColor,line width= 0.4pt,line join=round,line cap=round] ( 49.20, 66.48) -- ( 43.20, 66.48);

\path[draw=drawColor,line width= 0.4pt,line join=round,line cap=round] ( 49.20, 92.88) -- ( 43.20, 92.88);

\path[draw=drawColor,line width= 0.4pt,line join=round,line cap=round] ( 49.20,119.27) -- ( 43.20,119.27);

\path[draw=drawColor,line width= 0.4pt,line join=round,line cap=round] ( 49.20,145.67) -- ( 43.20,145.67);

\path[draw=drawColor,line width= 0.4pt,line join=round,line cap=round] ( 49.20,172.07) -- ( 43.20,172.07);

\path[draw=drawColor,line width= 0.4pt,line join=round,line cap=round] ( 49.20,198.47) -- ( 43.20,198.47);

\node[text=drawColor,rotate= 90.00,anchor=base,inner sep=0pt, outer sep=0pt, scale=  1.00] at ( 34.80, 66.48) {0.0};

\node[text=drawColor,rotate= 90.00,anchor=base,inner sep=0pt, outer sep=0pt, scale=  1.00] at ( 34.80, 92.88) {0.2};

\node[text=drawColor,rotate= 90.00,anchor=base,inner sep=0pt, outer sep=0pt, scale=  1.00] at ( 34.80,119.27) {0.4};

\node[text=drawColor,rotate= 90.00,anchor=base,inner sep=0pt, outer sep=0pt, scale=  1.00] at ( 34.80,145.67) {0.6};

\node[text=drawColor,rotate= 90.00,anchor=base,inner sep=0pt, outer sep=0pt, scale=  1.00] at ( 34.80,172.07) {0.8};

\node[text=drawColor,rotate= 90.00,anchor=base,inner sep=0pt, outer sep=0pt, scale=  1.00] at ( 34.80,198.47) {1.0};

\node[text=drawColor,anchor=base,inner sep=0pt, outer sep=0pt, scale=  1.00] at ( 42.80,69) {$\alpha$};

\path[draw=drawColor,line width= 0.4pt,line join=round,line cap=round] ( 49.20, 61.20) --
	(227.75, 61.20) --
	(227.75,203.75) --
	( 49.20,203.75) --
	( 49.20, 61.20);
\end{scope}
\begin{scope}
\path[clip] (  0.00,  0.00) rectangle (252.94,252.94);
\definecolor{drawColor}{RGB}{0,0,0}

\node[text=drawColor,anchor=base,inner sep=0pt, outer sep=0pt, scale=  1.00] at (138.47, 15.60) {$\mu/\|\epsi\|_2$};

\node[text=drawColor,rotate= 90.00,anchor=base,inner sep=0pt, outer sep=0pt, scale=  1.00] at ( 10.80,132.47) {power};
\end{scope}
\begin{scope}
\path[clip] ( 49.20, 61.20) rectangle (227.75,203.75);
\definecolor{drawColor}{RGB}{0,0,0}

\path[draw=drawColor,line width= 0.4pt,line join=round,line cap=round] ( 55.81, 70.81) --
	( 57.78, 70.98) --
	( 59.75, 71.65) --
	( 61.72, 72.58) --
	( 63.69, 73.80) --
	( 65.65, 75.59) --
	( 67.62, 77.83) --
	( 69.59, 80.21) --
	( 71.56, 83.28) --
	( 73.53, 86.52) --
	( 75.49, 90.79) --
	( 77.46, 95.13) --
	( 79.43,100.18) --
	( 81.40,105.84) --
	( 83.37,111.53) --
	( 85.33,117.68) --
	( 87.30,124.23) --
	( 89.27,130.66) --
	( 91.24,137.57) --
	( 93.21,144.18) --
	( 95.17,150.69) --
	( 97.14,156.83) --
	( 99.11,163.02) --
	(101.08,168.44) --
	(103.05,173.58) --
	(105.01,178.29) --
	(106.98,182.58) --
	(108.95,185.88) --
	(110.92,188.66) --
	(112.89,191.24) --
	(114.86,193.03) --
	(116.82,194.86) --
	(118.79,195.85) --
	(120.76,196.78) --
	(122.73,197.35) --
	(124.70,197.77) --
	(126.66,198.06) --
	(128.63,198.24) --
	(130.60,198.34) --
	(132.57,198.39) --
	(134.54,198.44) --
	(136.50,198.46) --
	(138.47,198.46) --
	(140.44,198.46) --
	(142.41,198.47) --
	(144.38,198.47) --
	(146.34,198.47) --
	(148.31,198.46) --
	(150.28,198.47) --
	(152.25,198.47) --
	(154.22,198.47) --
	(156.19,198.47) --
	(158.15,198.47) --
	(160.12,198.47) --
	(162.09,198.47) --
	(164.06,198.47) --
	(166.03,198.47) --
	(167.99,198.47) --
	(169.96,198.47) --
	(171.93,198.47) --
	(173.90,198.47) --
	(175.87,198.47) --
	(177.83,198.47) --
	(179.80,198.47) --
	(181.77,198.47) --
	(183.74,198.47) --
	(185.71,198.47) --
	(187.67,198.47) --
	(189.64,198.47) --
	(191.61,198.47) --
	(193.58,198.47) --
	(195.55,198.47) --
	(197.52,198.47) --
	(199.48,198.47) --
	(201.45,198.47) --
	(203.42,198.47) --
	(205.39,198.47) --
	(207.36,198.47) --
	(209.32,198.47) --
	(211.29,198.47) --
	(213.26,198.47) --
	(215.23,198.47) --
	(217.20,198.47) --
	(219.16,198.47) --
	(221.13,198.47);
\definecolor{drawColor}{RGB}{255,0,0}

\path[draw=drawColor,line width= 0.4pt,dash pattern=on 4pt off 4pt ,line join=round,line cap=round] ( 55.81, 70.89) --
	( 57.78, 71.02) --
	( 59.75, 71.61) --
	( 61.72, 72.30) --
	( 63.69, 73.62) --
	( 65.65, 75.19) --
	( 67.62, 77.22) --
	( 69.59, 79.30) --
	( 71.56, 82.42) --
	( 73.53, 85.24) --
	( 75.49, 89.21) --
	( 77.46, 93.11) --
	( 79.43, 97.62) --
	( 81.40,102.66) --
	( 83.37,108.14) --
	( 85.33,113.74) --
	( 87.30,119.64) --
	( 89.27,125.72) --
	( 91.24,131.90) --
	( 93.21,138.56) --
	( 95.17,144.19) --
	( 97.14,150.47) --
	( 99.11,156.57) --
	(101.08,162.19) --
	(103.05,167.35) --
	(105.01,171.87) --
	(106.98,176.37) --
	(108.95,179.98) --
	(110.92,183.58) --
	(112.89,186.59) --
	(114.86,188.81) --
	(116.82,191.25) --
	(118.79,192.84) --
	(120.76,194.43) --
	(122.73,195.39) --
	(124.70,196.13) --
	(126.66,196.89) --
	(128.63,197.32) --
	(130.60,197.67) --
	(132.57,197.92) --
	(134.54,198.07) --
	(136.50,198.19) --
	(138.47,198.28) --
	(140.44,198.33) --
	(142.41,198.38) --
	(144.38,198.42) --
	(146.34,198.44) --
	(148.31,198.44) --
	(150.28,198.44) --
	(152.25,198.45) --
	(154.22,198.46) --
	(156.19,198.46) --
	(158.15,198.46) --
	(160.12,198.46) --
	(162.09,198.47) --
	(164.06,198.46) --
	(166.03,198.47) --
	(167.99,198.47) --
	(169.96,198.46) --
	(171.93,198.47) --
	(173.90,198.47) --
	(175.87,198.47) --
	(177.83,198.47) --
	(179.80,198.47) --
	(181.77,198.47) --
	(183.74,198.47) --
	(185.71,198.47) --
	(187.67,198.47) --
	(189.64,198.47) --
	(191.61,198.47) --
	(193.58,198.47) --
	(195.55,198.47) --
	(197.52,198.47) --
	(199.48,198.47) --
	(201.45,198.47) --
	(203.42,198.47) --
	(205.39,198.47) --
	(207.36,198.47) --
	(209.32,198.47) --
	(211.29,198.47) --
	(213.26,198.47) --
	(215.23,198.47) --
	(217.20,198.47) --
	(219.16,198.47) --
	(221.13,198.47);
\definecolor{drawColor}{RGB}{0,0,0}

\path[draw=drawColor,line width= 0.4pt,dash pattern=on 1pt off 3pt ,line join=round,line cap=round] ( 49.20, 70.88) -- (227.75, 70.88);
\end{scope}
\end{tikzpicture}
		        \caption{Power for $n = 30$, with $\alpha = 1/30$.} \label{fig:D}
		    \end{subfigure}
		    
			\caption{Power plots comparing $\phi_{1/M}^{\calS}$ (solid line) to $\phi_{1/M}^{\calH_M}$ (striped line), where $\calS$ is an oracle subgroup of $\calH$, with $|\calS| = M$. Notice that the power of $\phi_{1/M}^{\calS}$ hits 1 exactly at $\sqrt{2}$, as predicted by Theorem \ref{thm:root2-tech}. Except for the $n = 2$ case, the power of the subgroup-invariance test dominates that of the Monte Carlo tests. In Appendix \ref{sec:conj}, we conjecture that this holds for all $n \geq 3$.} \label{fig:power}
		\end{figure}

		As an illustration, Figure \ref{fig:power} plots the power of a subgroup-invariance test based on a subgroup $\calS$ with $\delta_{\calS} = 0$, and a Monte Carlo group-invariance test based on $M$ draws from $\calH$, for $\calH$-invariant $\epsi$.
		Here, we indeed observe that the power of the subgroup-invariance test hits 1 at a signal-to-noise ratio of $\sqrt{2}$, as predicted by substituting $\delta_{\calS} = 0$ in Theorem \ref{thm:root2-tech}.
		This is especially visible in Figures \ref{fig:A} and \ref{fig:B}.
		At the same time, the power of Monte Carlo test only approaches 1 as $\mu / \|\epsi\|_2$ increases.
		 
	\subsection{Power comparison under normality}\label{sec:parametric}
		With the results discussed in the previous section, we obtain a power comparison in terms of consistency.
		To understand the power beyond consistency, we compare the two tests under normality.
		We connect Monte Carlo group-invariance tests to the $t$-test and, under normality, subgroup-invariance tests to the $Z$-test.
		This yields clean results about power of the tests, and allows for a simple comparison in the normal location model.
				
		The following result shows that the $t$-test can be interpreted as a group-invariance test.
		
		\begin{thm}\label{thm:t-test}
			The $t$-test is the orthogonal group-invariance test.
			That is, let $\widehat{\sigma} = \sqrt{X'(I - \iota\iota')X/(n-1)}$, then 
			\begin{align*}
				\phi_{\alpha}^{\calH}(X) = 
					\mathbb{I}{\{\iota'X / \widehat{\sigma} > t_{n-1}^{\alpha}\}},
			\end{align*}
			where $t_{n-1}^{\alpha}$ denotes the $\alpha$ upper-quantile of the $t$-distribution with $(n-1)$ degrees of freedom.
		\end{thm} 

		\begin{rmk}
			Note that this result does not depend on the distribution of $X$, so for this result we do not assume $X$ is normally distributed.
			In fact, it holds conditional on $X$.
		\end{rmk}
		
		We do not believe Theorem \ref{thm:t-test} is novel. 
		However, to our surprise, we were unable to find the result in the literature or textbooks in this form, although several strongly related results exist: see \citet{chmielewski1981elliptically}.
		For example, the result does not appear in \citet{lehmann2005testing}, who extensively discuss group-invariance tests and their relationship to the $t$-test (see Chapter 15.2 and in particular Example 15.2.4).
		The result is straightforward to generalize to $F$-tests to test parameters of higher dimension.
		
		The proof of Theorem \ref{thm:t-test} can be modified to obtain an analogous result for Monte Carlo group-invariance tests.
		
		\begin{cor}\label{cor:MC_t}
			The test $\phi_{\alpha}^{\calH_{M}}$ has the same size and power as an $M$-draw Monte Carlo $t_{n-1}$-test.
		\end{cor} 	

		For subgroups with $\delta_{\calS}^{\text{abs}} = 0$, we establish a similar connection to the $Z$-test.
		
		\begin{thm}\label{thm:norm}
			Let $\calS$ be a subgroup of $\calH$ with $\delta_{\calS}^{\text{abs}} = 0$.
			Let $\epsi \sim \calN(0, \sigma^2I)$, where $\sigma^2 > 0$.
			Then, an $\calS$-invariance test has the same size and power as an $|\calS|$-Monte Carlo $Z$-test.
		\end{thm} 
	
		Intuitively, the result states that if $\epsi$ happens to have distribution $\epsi \sim N(0, \sigma^2I)$, $\sigma^2 > 0$, unbeknownst to the analyst, then the $\calS$-invariance test has the same size and power as a Monte Carlo $Z$-test.
		The subgroup therefore essentially allows the analyst to sample from the unknown null distribution.
		For this reason, we henceforth refer to subgroups $\calS$ with $\delta_{\calS}^{\text{abs}} = 0$ as \textit{oracle} subgroups.
		In Section \ref{sec:oracle} we discuss the existence and order of these subgroups.

		Comparing Theorem \ref{thm:norm} to Corollary \ref{cor:MC_t}, we conclude that oracle subgroup-invariance tests are more powerful than MC group-invariance tests in a normal location model, as captured in
		\begin{cor}\label{cor:oracle_vs_sampling}
			Let $\vepsi \sim \calN(0, \sigma^2I)$, $\sigma^2 > 0$ and $\calS \subset \calH$ be a subgroup for which $\delta_{\calS}^{\text{abs}} = 0$.
			Then, an $\calS$-invariance test is more powerful than an $|\calS|$-Monte Carlo $\calH$-invariance test.
		\end{cor}

\section{Generalized location model: choosing a subgroup and group codes}\label{sec:group_codes}
		In this section, we consider some important properties of subgroups $\calS$ of $\calH$ for which $\delta_{\calS}$ is `small'.		
		First, notice that $-1 \leq \delta_{\calS} \leq 1$ for every $\calS$, so that it makes sense to focus on subgroups for which $\delta_{\calS} < 1$.
		Let us write $\calS\iota := \{S\iota\ |\ S \in \calS\}$.
		For such subgroups, we then have
		
		\begin{prp}\label{prp:low_leak}
			If $\delta_{\calS} < 1$, then $\calS$ is finite and the map $\calS \mapsto \calS\iota$ is bijective.
		\end{prp}

		This result allows us to represent a subgroup by a matrix $\fS$ with columns $S\iota$, $S \in \calS$.
		These columns can be interpreted as the rotations of $\iota$ by elements of $\calS$.
		This matrix representation of the subgroup yields the following representation of $\delta_{\calS}$.
		
		\begin{prp}\label{prp:leak_representation}
			$\delta_{\calS}	= \max_{i \neq j} e_i'\fS'\fS e_j$.
		\end{prp}
		
		Notice that $\fS'\fS$ contains all the inner-products between the columns of $\fS$, which are all $n$-dimensional unit vectors and so are located on the unit hypersphere in dimension $n$.
		The value $\delta_{\calS}$ can therefore be interpreted as the maximum inner-product between two of such points on the unit hypersphere, which has a one-to-one correspondence with the minimum angle, $\arccos(\delta_{\calS})$, between any two points.
		Hence, the problem of minimizing $\delta_{\calS}$ is equivalent to finding a group that induces points on a hypersphere that are as far away from each other as possible.
		Such a collection of points on a hypersphere is also known as a \emph{group code} \citep{slepian1968group, conway1998sphere}.
		The problem of finding $M$, say, such points on a hypersphere such that $\delta_{\calS}$ is minimized is a group-restricted version of the so-called \emph{Tammes problem}.
		
		It is well-established that a moderately large value of $M$ often yields a small minimum value of $\delta_{\calS}$.
		For example, \citet{sloane1996spherical} lists a group code for $n = 16$ and $M = 256$ that is induced by a subgroup of the sign-flipping group $\calR$, for which $\delta_{\calS} = .25$.
		As a comparison, we find an average `$\delta$' $\approx .68$ from $10^5$ random subsets of size $M$ from $\calR$.
		This suggests that a carefully chosen subgroup should typically be able to yield a much smaller leak than a Monte Carlo draw.
		
		\begin{rmk}	
			Unfortunately, we were unable to find suitable algorithms to construct `good' group codes of order $M$, nor could we find libraries that contain them.
			For example, \citet{sloane1996spherical} lists spherical codes up to $n = 24$ for few values of $M$, only some of which are group codes.
			In addition, we may not be satisfied with any `good' group code: the group code needs to be induced by a subgroup of the group under which invariance is assumed.
			Therefore, we include a simple algorithm in Appendix \ref{appn:alg} for the case of the sign-flipping group.
		\end{rmk}

		\begin{rmk}
			Notice that $\delta_{\calS}^{\text{abs}}$ and $\delta_{\calS}$ are connected through the identity $\delta_{\calS}^{\text{abs}} = \delta_{\calS \cup (-\calS)}$, where $-\calS := \{-S\ |\ S \in \calS\}$.
			Hence, we can also view $\delta_{\calS}^{\text{abs}}$ as a minumum angle between points on a hypersphere, noting that each point is accompanied by a twin on the other side of the hypersphere.
			Alternatively, $\delta_{\calS}^{\text{abs}}$ can be interpreted as a minimum angle between $|\calS|$ lines that pass the origin and an element in $\calS\iota$.
		\end{rmk}

		\subsection{Oracle subgroups}\label{sec:oracle}
			An important special case are subgroups for which $\delta_{\calS}^{\text{abs}} = 0$.
			Such subgroups are `optimal' for two-sided testing in the sense of Theorem \ref{thm:root2-tech-abs}.
			We will refer to such subgroups as \emph{oracle subgroups}, due to their power properties discussed in Section \ref{sec:parametric}.
			From Propositions \ref{prp:low_leak} and \ref{prp:leak_representation}, we know that such subgroups are represented by $n$-row orthonormal matrices: all the off-diagonals of $\fS'\fS$ are zero.
			Combining this with the Gram-Schmidt Theorem proves
			\begin{prp}\label{prp:oracle_n}
				The maximum order of an oracle subgroup $\calS$ is $n$.	
			\end{prp}
			
			Furthermore, we include a result about the existence of oracle subgroups.
			\begin{prp}\label{prp:existence}
				There exists an oracle subgroup $\calS$ of $\calH$ with respect to any unit vector $\viota$, of any order $p$, with $1 \leq p \leq n$.
			\end{prp} 
		
			Unfortunately, even for quite `large' subgroups $\calG$ of $\calH$, there often exist values $p \leq n$ such that $\calG$ has no oracle subgroups of order $p$.
			In addition, the existence of oracle subgroups of $\calG$ of $\calH$ depends intimately on the choice of $\viota$.
			This will be seen in Section \ref{sec:sign-flipping}, where we characterize oracle subgroups of the sign-flipping group for $\viota = n^{-1/2}(1, 1, \dots, 1)'$.		

	\section{Example: sign-flipping}\label{sec:sign-flipping}
	Up to this point, we have only discussed subgroups in an abstract sense.
	In this section, we provide some examples by considering the sign-flipping group $\calR$, which can be represented by all diagonal matrices with diagonal elements in $\{-1, 1\}$ under matrix multiplication.
	In addition, we choose $\iota = n^{-1/2}(1, 1, \dots, 1)'$.
	
	If $\epsi$ is $\calR$-invariant and we additionally assume that its elements are independent, then the elements of $\vepsi$ are marginally symmetrically distributed about the origin.
	The resulting test is often used in paired data, as was already proposed by \citet{fisher1935}. 
	There, the idea is to sign-flip differences between paired observations.
	Sign-flipping is also widely used in brain image analyses, see Section \ref{sec:application}.
	For additional discussions and applications, see \citealt{efron1969student,bekker2008symmetry,davidson2008wild,winkler2014permutation,andreella2020permutation}.
	
	In order to study the leak of subgroups of $\calR$, it is convenient to use its matrix representation $\fR := (\iota, R_1\iota, R_2\iota, \dots)$, $R_1, R_2, \dots \in \calR$.
	The columns of $n^{1/2}\fR$ are the diagonals of sign-flipping matrices in $\calR$.
	The same holds for a subgroup $\calS \subset \calR$ and its analogous matrix representation $\fS$.
	In fact, the group structure is preserved by considering element-wise multiplication of the columns of $n^{1/2}\fS$.	
	So, the matrix $n^{1/2}\fS$ fully describes the subgroup.
	
	Notice that the distribution of $n\iota'\overline{S}\iota$, where $\overline{S}$ is uniformly distributed on $\calS$, coincides with the empirical distribution over $n\iota'\fS$.
	We refer to this distribution as the `leak distribution'.
		
	\begin{exm}\label{exm:n=2}
		If $n = 2$, then 
		\begin{align*}
			n^{1/2}\fR = 
			\begin{bmatrix}
				1 & 1  & -1 & -1 \\
				1 & -1 & 1  & -1
			\end{bmatrix}.
		\end{align*}
		The leak distribution of $\calR$ is then the empirical distribution over
		\begin{align*}
			n \iota'\fR =
			\begin{bmatrix}
				2 & 0 & 0 & -2
			\end{bmatrix},
		\end{align*}
		which assigns .25 mass to both 2 and -2, and .5 mass to 0.
		Notice that if $n = 2$, then $\delta_{\calR} = 0$.
		However, $\delta_{\calR}^{\text{abs}} = 2$, so that $\calR$ is not an `oracle' subgroup.
	\end{exm}

	\begin{exm}\label{exm:n=2_oracle}
		If $n = 2$, then an example of an oracle subgroup $\calS$ of $\calR$ is 
		\begin{align*}
			n^{1/2}\fS = 
			\begin{bmatrix}
				1 & 1  \\
				1 & -1
			\end{bmatrix}.
		\end{align*}
		The leak distribution is uniform on
		\begin{align*}
			n\iota'\fS =
			\begin{bmatrix}
				1 & 0 
			\end{bmatrix}.
		\end{align*}
		Here, both $\delta_{\calS} = 0$ and $\delta_{\calS}^{\text{abs}} = 0$.
	\end{exm}	
	
	\subsection{Subgroups of the sign-flipping group}
		In this section, we describe some (well-known) properties of the sign-flipping group, as well as the induced leak distributions.
		As $\calR$ is isomorphic to a boolean group, its subgroups are of order $2^p$, for some $p \leq n$, $p \in \mathbb{N}$, where $p$ the called the rank of the subgroup.
		The subgroups of the sign-flipping group are abundant, even if $n$ is small.
		The number of subgroups of rank $p$ is equal to the $p$th element of the $n$th row of the 2-binomial coefficient triangle listed as entry A022166 in the \citet{oeisA022166}.
		The total number of subgroups is equal to the sum of the $n$th row of this triangle, which can be found in entry A006116 of the \citet{oeisA006116}.
		This means that if $n = 9$, say, then we have 3309747 subgroups of rank $p = 4$, and 8283458 subgroups in total. 
		
		While the number of different subgroups is large, many of them yield the same vector $n^{1/2}\iota'\fR$, and hence the same leak distribution.
		In particular, the number of different leak distributions corresponding to a subgroup of rank $p$ is equal to the $p$th element of the $n$th row of the triangle listed as entry A076831 of the \citet{oeisA076831}.
		The total number of different distributions is equal to the sum of the $n$th row of this triangle, which can be found in entry A076766 of the \citet{oeisA076766}.
		For example, for $n = 9$ there are 240 different leak distributions corresponding to subgroups of rank $p = 4$, and 848 different leak distributions in total.
		
		\begin{exm}
			For $n = 4$ and subgroups of order 4 (so $p = 2$), there exist 6 different leak distributions.
			These distributions are illustrated in Figure \ref{fig:leak_distr}.
			As we can see in the figure, the leak distributions are quite diverse.
			The fifth image corresponds to an oracle subgroup: except for the identity element, all mass is at 0 so that $\delta_{\calS}^{\text{abs}} = 0$.
			The second, third and fifth image all have $\delta_{\calS} = 0$.
			\begin{figure}
				\includegraphics[width = 1\linewidth]{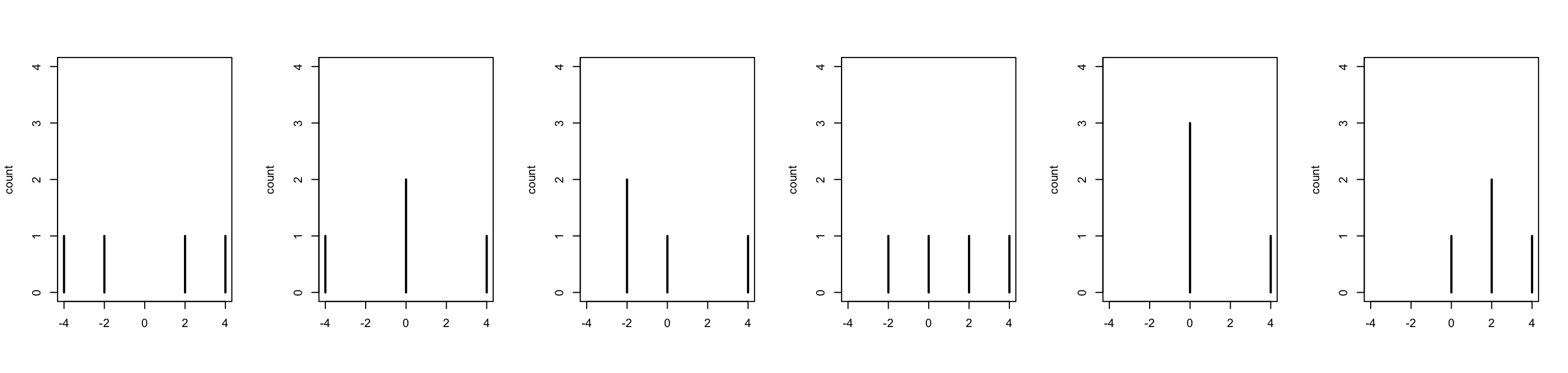}
				\caption{Histograms of leak distributions for all subgroups of $\calR$ of order $4$ for $n = 4$. The fifth histogram corresponds to an oracle subgroup, as all its mass it at zero, except for the mass at $n$ produced by the identity element. Notice that the leak distributions are quite diverse.}
				\label{fig:leak_distr}
			\end{figure}
		\end{exm}
		
	\subsection{Oracle and near-oracle subgroups}\label{sec:sign-oracle}
		The structure of the sign-flipping group $\calR$ and choice $\iota = (n^{-1/2}, n^{-1/2}, \dots)$ allow us to easily characterize the order of the oracle subgroups.
		See the appendix for a constructive proof.
				
		\begin{thm}\label{thm:2primefact}
			Let $k \leq l$, $k \in \mathbb{N}_+$, where $l$ is the number of $2$s in the prime factorization of $n$.
			Then $\calR$ has an oracle subgroup with respect to $\iota = (n^{-1/2}, n^{-1/2}, \dots)$ of order $2^k$.
			Furthermore, if $\calS$ is an oracle subgroup of $\calR$, then it is of order $2^k$ for some $k \in \mathbb{N}_+$.
		\end{thm}
		
		Theorem \ref{thm:2primefact} implies that the number of 2's in the prime factorization of $n$ determines the maximum cardinality of its oracle subgroups.
		In particular if $n = 2^k$, for some $k \in \mathbb{N}$, then there exists an oracle subgroup of order $n$, which is the largest order that exists by Proposition \ref{prp:oracle_n}.
		However, if $n$ is an odd number, then the only oracle subgroup of $\calR$ that exists is the trivial subgroup containing only the identity element.
		In Appendix \ref{appn:alg} we include a simple algorithm to compute subgroups $\calS$ of $\calR$ for which $\delta_{\calS}$ or $\delta_{\calS}^{\text{abs}}$ is small.

	\section{Application: fMRI data}\label{sec:application}
	In this article we have mainly focused on the problem of testing a single hypothesis. However, the idea of using near-oracle subgroups directly extends to permutation-based multiple testing methods (\citealp{westfall1993resampling, tusher2001significance, meinshausen2006false, pesarin2010permutation, meinshausen2011asymptotic, hemerik2018false, blanchard2020post}). Such methods allow testing a large number of hypotheses simultaneously. A main advantage of permutation-based multiple testing method as compared to other multiple testing methods, is that they take into account the dependence structure in the data, leading to good power properties \citep{westfall1993resampling, hemerik2018false, hemerik2019permutation}. Like permutation tests, these methods require using a group of transformations or  transformations randomly sampled from a group. In the present data analysis example, we will use a near-oracle subgroup within a permutation-based multiple testing method from \citet{hemerik2018false}.

	Permutation-based multiple testing methods can be computationally demanding. The first reason is that for each permutation we need to compute a large number of test statistics, equal to the total number of hypotheses. The second reason is that some multiple testing methods apply sophisticated combinatorical algorithms to the matrix of computed test statistics.

	The multiple testing method that we will use here is the \emph{approximate closed testing method} from \citet[][p.144]{hemerik2018false}. An implementation has been available in \citet{Hemerik_confSAM_2018}. Recently a faster implementation (of the closed testing method) has become available, which we use here, see \citet{Koning_fast_confSAM_2022}.

	We applied the multiple testing metod to a high-dimensional functional magnetic resonance imaging (fMRI) dataset. 
	The original data are available at \url{https://openfmri.org} and we used the pre-processed data from \citet{Andreella_fMRIdata_2021} (for details see \citealp{smeets2013allured, andreella2020permutation}). 
	The dataset contains measurements for $n=29$ subjects who interchangeably looked a images of food and non-food. 
	Thus, the subjects were exposed to two experimental conditions. 
	For 152472 voxels making up the brain, the  activity was recored while  the subjects looked at the images. For each subject and for voxel $i$, a difference statistic was computed with mean $\mu_i$ say.
	 For voxel $i$, we define the corresponding null hypothesis to be $H_i: \mu_i=0$, which mean that there is no difference in mean response between the two conditions.
 
	 For each voxel  we computed a t-statistic based on the 29 measurements. 
	 We assumed that under $H_i$, the correponding t-statistic was symmetric about 0. 
	 This framework allows us to use the permutation-based multiple testing method from \citet{hemerik2018false}, where rather than permutation, we  used sign-flipping (we took $|\mathcal{S}|=10^3$, see \citealp{hemerik2018false}).  
	 The method requires the user to set a rejection threshold. 
	 All hypotheses with test statistics exceeding this threshold are rejected. 
	 For our illustration purposes, we simply set the the threshold to 3, so that all hypotheses with t-statistic above 3 or below $-3$ were rejected.
	 This led to 10580 rejected hypotheses, i.e. 10580 voxels were selected as seemingly `activated'.
 
	 The multiple testing method provides a (median unbiased) estimate and a $95\%$-confidence upper bound for the \emph{false discovery proportion} (FDP), which is the fraction of incorrect rejections among all 10580 rejected null hypotheses \citep{hemerik2018false}. 
	 As our simulations show (see Section \ref{sec:simulations}), by using a near-oracle subgroup within a sign-flipping test, we only require about half the number of sign-flips compared to using random sign-flipping, to achieve the same power. 
	 This extends to permutation-based multiple testing methods, which are based on the same group invariance principle.

	This means that where one would use perhaps 2000 random transformations, we can instead use a near-oracle subgroup of cardinality 1024 and likely have comparable power, as well as improved replicability.
	We did the latter in this example. 
	Note that this reduces the computation time with about $50\%$ on average, compared to using 2000 random sign-flips.
	We conveniently obtained the sign-flipping matrix encoding the near-oracle subgroup from \citet{Koning_Database_of_Near_2022}.
	Thus, it was straightforward to implement the near-oracle subgroup within the multiple testing method.

	In our example the computation time on a laptop was 6 minutes, counting from the moment that the multiple testing method was called. 
	The method  estimated the FDP to be $360/10580\approx .034$ and provided a $95\%$-confidence upper bound for the FDP of $2306/10580\approx .217$.
	This means that we can be confident that most of the 10580 selected voxels are truly activated, i.e. that their activity depends on the experimental condition (looking at food vs. looking at non-food).
	Interpreting the results further is beyond the scope of this paper.
	Our analysis illustrates that near-oracle subgroups can easily be used within permutation-based multiple testing procedures.

		\section{Simulation results}\label{sec:simulations}
		In this section, we present some simulation results.
		We simulated data $X$ using the standard normal location model
		\begin{align*}
			X = \iota\mu + \epsi, \quad \epsi \sim \calN(0_n, I_n),
		\end{align*}
		with $\iota = (n^{-1/2}, n^{-1/2}, \dots)'$.
		We tested the hypothesis $H_0: \mu = 0$ against $H_1 : \mu > 0$, using the sign-flipping invariance assumption $\vepsi \eqd R\vepsi$, $R \in \calR$. 
		Notice that this assumption indeed holds for the standard normal location model, so that $\epsi$ is $\calR$-invariant.
		
		We used the following tests, where the abbreviation between brackets corresponds to the column names in the simulation tables.
		The tests are grouped by the number of elements in the random subset or subgroup, both indicated by $M$.
		\begin{itemize}
			\item Two benchmark tests to provide upper bounds on the power. In particular, since they are based on supergroups of the subgroups we consider, their power is an upper bound power of the subgroup-invariance tests, by Theorem \ref{thm:subgroup_vs_full_group} and Theorem \ref{thm:t-test}.
			\begin{itemize}
				\item A $t$-test, which exploits knowledge about the orthogonal invariance of the distribution of $\vepsi$ ($t$).
				\item A group-invariance test based on $\calR$ (if computationally feasible) or a Monte Carlo sign-flipping test based on 1000 draws ($\calR$ or MC $\calR$).
			\end{itemize}
			\item Tests based on $M = n$ draws:
			\begin{itemize}
				\item An oracle subgroup-invariance test (Oracle).
				\item An $n$-draw Monte Carlo $Z$-test (MC $Z$).
				\item An order $n$ subgroup-invariance test based on a subgroup with $\delta_{\calS}= 0$ and $\iota'S\iota < 0$ for some $S \in \calS$ (Neg.).
				\item An $n$-draw Monte Carlo $\calR$-invariance test (MC $\calR$).
			\end{itemize}
			\item Tests with $M = 2n$:
			\begin{itemize}
				\item An order $2n$ subgroup-invariance test based on a subgroup with $\delta_{\calS}= 0$ and $\iota'S\iota < 0$ for some $S \in \calS$ (Neg.).
				\item A $2n$-Monte Carlo $\calR$-invariance test (MC $\calR$).
			\end{itemize}
			\item Tests with $M = 4n$:
			\begin{itemize}
				\item A subgroup-invariance test based on a subgroup for which $\delta_{\calS}$ is `small', produced by Algorithm \ref{alg:near-oracle} described in Appendix \ref{appn:alg} (NOS).
				\item A $4n$-Monte Carlo $\calR$-invariance test (MC $\calR$).
			\end{itemize}
		\end{itemize}
		
		We considered $n \in \{8, 16, 32, 64, 128\}$, to guarantee the existence of oracle subgroups of $\calR$ of order $n$, by Theorem \ref{thm:2primefact}. 
		We chose the level $\alpha$ such that $\alpha n$ is integer and $\alpha \approx .05$.
		This ensures all tests considered have size $\alpha$ \citep{hemerik2018exact}.
		The parameter $\mu$ was chosen such that the power is sufficiently far away from $\alpha$ and $1$.
		For each setting and test, we generated $X$ $10^6$ times, independently across iterations, tests and settings, and recorded the proportion of times the tests rejected the null hypothesis. 
		The results are reported in Tables \ref{tab:n=8} to \ref{tab:n=128}.
		
		Our findings are as follows.
		As expected, the rejection proportion under $H_0$ is approximately $\alpha$ for all tests, because all tests are exact.
		The benchmark $t$-test outperforms the other tests, which is unsurprising as it makes explicit use of orthogonal invariance, to which the other tests do not have access.
		The 1000-MC $\calR$ test and $\calR$-invariance tests perform are slightly less powerful than the $t$-test.
		In line with Theorem \ref{thm:norm}, the oracle subgroup-invariance tests have the same power as the MC $Z$-tests.
		Furthermore, they are outperformed by the `negative' subgroup tests, based on a subgroup with $\delta_{\calS}= 0$ and $\iota'S\iota < 0$ for some $S \in \calS$.
		
		We now turn to the comparison of the (sub)group-invariance tests and the MC group-invariance tests.
		All order $M$ subgroup-invariance tests we consider outperform the corresponding $M$-MC $\calR$ tests.
		This is especially the case if $M$ and $n$ are small, where we find a power gap of over 6 percentage points between the negative oracle subgroup-invariance and $n$-MC tests in the most extreme case where $\mu = .7$, $n = 8$.		

		Next to this absolute power comparison, we also compare the power in a relative sense.
		In particular, we consider the  relative power difference to the $t$-test, which serves as an upper bound of the subgroup tests and proxy for the $\calR$-invariance test.
		Here, we find that in all cases the power difference between the $M$-MC $\calR$ test to the $t$-test is 1.5 to 5 times larger than power difference between the best order $M$ subgroup-invariance test and the $t$-test.
		So while the absolute power difference seems small in some cases, this is only because the power of both tests is close to an upper bound: the relative power gap to the bound is substantial.
	
		Indeed, notice that as $M$ increases, both the MC sample (without replacement) from $\calR$ and subgroup converge to the entire group $\calR$. 
		Hence, the $M$-MC $\calR$ and order $M$ subgroup-invariance tests both converge towards the $\calR$-invariance test as $M$ increases.
		Therefore, we expect the power gap between the two tests to decrease as $M$ increases.
		This is corresponds to what is observed in the simulations.

		Another way of seeing this power gap, is by comparing across different values of $M$.
		Doing so, we find that in each case we consider, the power of the $M$-MC $\calR$ test is closer to that of the best order $M/2$ subgroup-invariance test than the power of the corresponding order $M$ subgroup-invariance test.
		Furthermore, in many cases the $M$-MC $\calR$ test is outperformed by the best order $M/2$ subgroup-invariance test.
		This means we could halve $M$ if we switch from an MC $\calR$ test to a subgroup-invariance test, and retain a similar power.
		
		Finally, we studied the variability (conditional on the data) of the $p$-value of the subgroup invariance test, as compared to the test based on random transformations.
		Of course, the $p$-value of the subgroup invariance test is completely deterministic given the data.
		However, it should be noted that simply reordering the data vector may lead to a different $p$-value.
		Hence, we also considered the effect of a random permutation of the data before applying the near-oracle subgroup-invariance test.
		We found that the $p$-value is still less variable than when random sign-flipping is used.
		As an example, for $n = 20$, $\mu = .5$, and $M = 64$ we found that on average (based 1000 datasets from a standard normal distribution and 1000 random permutations) the average variance of the $p$-value was only $.00028$, while the average variance was $.00071$ if $M$ random transformations were used.
		We obtained comparable results in other settings.
		For example, for $n = 32$, $\mu = .3$ and $M = 64$, the average variances were $.00038$ and $.00115$, respectively.
		
		\clearpage
		%

\begin{table}[!htb]
	\centering
	\scriptsize
	\hspace{-1.2cm}
	\begin{tabular}{ c | c | c || c | c | c | c|| c | c || c | c }
		& \multicolumn{2}{c||}{} & \multicolumn{4}{c||}{$M = n = 8$} & 
		\multicolumn{2}{c||}{$M = 2n = 16$} & \multicolumn{2}{c}{$M = 4n = 32$} \\
		\hline
		$\mu$ & $t$ & $\calR$ & Oracle & MC $Z$ & Neg. & MC $\calR$ & Neg. & MC $\calR$ & NOS & MC $\calR$ \\
		\hline
		.0 & .12587 & .12491 & .12483 & .12531 & .12511 & .12533 & .12463 & .12494 & .12460 & .12539 \\
		.3 & .38145 & .36486 & .33627 & .33643 & .34395 & .32267 & .36149 & .34346 & .36228 & .35488 \\
		.5 & .60173 & .57593 & .52532 & .52276 & .53296 & .49510 & .56675 & .53448 & .57301 & .55591 \\ 
		.7 & .79648 & .76700 & .70453 & .70425 & .71680 & .65678 & .75625 & .71144 & .76114 & .74068 \\ 
	\end{tabular}
	\caption{$n = 8$, $10^6$ simulations, $\alpha = 1/8 = .125$ and $|\calR| = 2^8 = 256$.}
	\label{tab:n=8}
\end{table}

		%

\begin{table}[!htb]
	\centering
	\scriptsize
	\hspace{-1.2cm}
	\begin{tabular}{ c | c || c || c | c | c | c || c | c || c | c}
		& \multicolumn{1}{c||}{} & $M = 1000$ & \multicolumn{4}{c||}{$M = n = 16$} & 
		\multicolumn{2}{c||}{$M = 2n = 32$} & \multicolumn{2}{c}{$M = 4n = 64$} \\
		\hline
		$\mu$ & $t$ & MC $\calR$ & Oracle & MC $Z$ & Neg. & MC $\calR$ & Neg. & MC $\calR$ & NOS & MC $\calR$ \\
		\hline
		.0 & .06223 & .06290 & .06261 & .06292 & .06221 & .06276 & .06282 & .06259 & .06274 & .06225 \\ 
		.3 & .36921 & .35172 & .32162 & .32090 & .32429 & .30469 & .34416 & .32563 & .34858 & .33771 \\ 
		.5 & .67925 &  .65130 & .59473 & .59520 & .59972 & .55808 & .63810 & .60127 & .64282 & .62578 \\ 
		.7 & .89667 & .87485 & .82502 & .82513 & .82940 & .77991 & .86511 & .83035 & .86929 & .85370 \\ 
	\end{tabular}
	\caption{$n = 16$, $10^6$ simulations, $\alpha = 1/16 = .0625$.}
	\label{tab:n=16}
\end{table}



\begin{table}[!htb]
	\centering
	\scriptsize
	\hspace{-1.2cm}
	\begin{tabular}{ c | c || c || c | c | c | c || c | c || c | c}
		& \multicolumn{1}{c||}{} & $M = 1000$ & \multicolumn{4}{c||}{$M = n = 32$} & 
		\multicolumn{2}{c||}{$M = 2n = 64$} & \multicolumn{2}{c}{$M = 4n = 128$} \\
		\hline
		$\mu$ & $t$ & MC $\calR$ & Oracle & MC $Z$ & Neg. & MC $\calR$ & Neg. & MC $\calR$ & NOS & MC $\calR$ \\
		\hline
		.0 & .06206 & .06302 & .06289 & .06257 & .06264 & .06280 & .06225 & .06251 & .06216 & .06299 \\ 
		.3 & .56387 & .55208 & .52535 & .52520 & .52789 & .51110 & .54648 & .53152 & .54867 & .54139 \\ 
		.4 & .76734 & .75342 & .72365 & .72298 & .72674 & .70666 & .74676 & .72930 & .74997 & .74191 \\ 
		.5 & .90212 & .89239 & .86899 & .86932 & .87126 & .85343 & .88734 & .87414 & .89016 & .88319 \\ 
	\end{tabular}
	\caption{$n = 32$. $10^6$ simulations. $\alpha = 2/32 = .0625$.}
	\label{tab:n=32}
\end{table}


\begin{table}[!htb]
	\centering
	\scriptsize
	\hspace{-1.2cm}
	\begin{tabular}{ c | c || c || c | c | c | c || c | c || c | c}
		& \multicolumn{1}{c||}{} & $M = 1000$ & \multicolumn{4}{c||}{$M = n = 64$} & 
		\multicolumn{2}{c||}{$M = 2n = 128$} & \multicolumn{2}{c}{$M = 4n = 256$} \\
		\hline
		$\mu$ & $t$ & MC $\calR$ & Oracle & MC $Z$ & Neg. & MC $\calR$ & Neg. & MC $\calR$ & NOS & MC $\calR$ \\
		\hline
		.0 & .04695 & .04687 & .04680 & .04681 & .04675 & .04689 & .04675 & .04668 & .04703 & .04720 \\ 
		.2 & .46994 & .46191 & .44659 & .44765 & .44811 & .43941 & .45895 & .45183 & .46121 & .45681 \\ 
		.3 & .76499 & .75592 & .73785 & .73748 & .73901 & .72858 & .75187 & .74250 & .75503 & .75005 \\ 
		.4 & .93598 & .93132 & .91993 & .92019 & .92045 & .91379 & .92932 & .92312 & .92988 & .92818 \\ 
	\end{tabular}
	\caption{$n = 64$. $10^6$ simulations. $\alpha = 3/64 = .046875$.}
	\label{tab:n=64}
\end{table}

		\begin{table}[!htb]
	\centering
	\scriptsize
	\hspace{-1.2cm}
	\begin{tabular}{ c | c || c || c | c | c | c || c | c || c | c}
		& \multicolumn{1}{c||}{} & $M = 1000$ & \multicolumn{4}{c||}{$M = n = 128$} & 
		\multicolumn{2}{c||}{$M = 2n = 256$} & \multicolumn{2}{c}{$M = 4n = 512$} \\
		\hline
		$\mu$ & $t$ & MC $\calR$ & Oracle & MC $Z$ & Neg. & MC $\calR$ & Neg. & MC $\calR$ & NOS & MC $\calR$ \\
		\hline
		.0 	& .04650 & .04695 & .04674 & .04668 & .04673 & .04651 & .04696 & .04661 & .04684 & .04695 \\ 
		.15 & .50894 & .50331 & .49668 & .49663 & .49662 & .49247 & .50319 & .49862 & .50365 & .50190 \\ 
		.2 	& .72134 & .71603 & .70804 & .70760 & .70884 & .70309 & .71548 & .70957 & .71620 & .71327 \\ 
		.25 & .87495 & .87163 & .86480 & .86444 & .86452 & .86116 & .87065 & .86611 & .87102 & .86942 \\ 
	\end{tabular}
	\caption{$n = 128$. $10^6$ simulations. $\alpha = 6/128 = .046875$.}
	\label{tab:n=128}
\end{table}

		%

	\section{Discussion}
	In this paper, we have proposed subgroup-invariance tests, which are based on a subgroup of the group under which invariance is tested.
	These tests are exact for testing invariance under the full group, and deterministic.
	In a generalized location model, we contrasted these tests with the popular Monte Carlo group-invariance tests, and compared their power.
	We found that subgroup-invariance tests tend to outperform their Monte Carlo counterparts, in the sense of extracting more power from the same number of transformations. 
	
	In practice, before our approach can be used, a requirement is the determination of an appropriate subgroup. 
	Fortunately, subgroups can be pre-computed, saved and then used indefinitely, as subgroup-invariance tests are exact conditional on the choice of subgroup.
	
	For example, our R package NOSdata \citep{Koning_Database_of_Near_2022} serves as a public source for downloading a `good' sign-flipping subgroup for $\iota = n^{-1/2}(1, 1, \dots, 1)'$ of appropriate dimensions. 
	This has many important applications, such as testing in linear models \citep{winkler2014permutation}, wild bootstrap methods \citep{davidson2008wild}, neuroimaging \citep{andreella2020permutation, blain2022notip}, and generalized linear models \citep{hemerik2020robust, hemerik2021permutation, desantis2022inference}.
	In addition, in Appendix \ref{appn:two_group}, we describe how sign-flipping subgroups can sometimes be transformed into subgroups for another setting.
	
	An important area for future research is the development of general algorithms to construct desired subgroups for testing invariance under other groups and for different $\iota$.
	Here, it is important to note that while it may be hard to find the subgroup that yields `optimal' power properties, it may be quite easy to find a subgroup that yields `good' power properties.		
	Our consistency and simulation results suggest that such a `good' subgroup $\calS$, for which $\delta_{\calS}$ is small but not minimal, is still expected to have good power properties.
	
	An application to testing exchangeability of binary sequences is discussed in the master's thesis of \citet{clemens2021enhancing}, which is based on an early version of this paper.
	In particular, they test exchangeability against an alternative that generates streaks of zeros and ones.
	They construct subgroups geared towards `breaking' such streaks, and find some subgroups that can indeed outperform Monte Carlo tests in a simulation study.
	\section{Data available statement}
	The original data for the application are available at \url{https://openfmri.org} and we used the pre-processed data from \citet{Andreella_fMRIdata_2021} (for details see \citealp{smeets2013allured, andreella2020permutation}).
	The code for the confSAM analysis is available in \citet{Koning_fast_confSAM_2022}.
	The subgroups used in the simulations can be found in \citet{Koning_Database_of_Near_2022}, and an implementation of the algorithm used to construct them in \citet{Koning_Near_Oracle_Subgroups_2022}.

\section{Funding statement}
	We have no funding to declare.
	\begin{appendix}
		\section{Two-sample comparison}\label{appn:two_group}
	In this section, we explain how the two-sample comparison problem fits into the location model described in Section \ref{sec:location-model}.
	In particular, we consider testing the equality of means $\mu_1$ and $\mu_2$ of two samples which we will denote by $X_1$ and $X_2$.
	The samples contain $m_1$ and $m_2$ observations respectively, where $m_1 + m_2 = n$.
	Define $1_m = n^{-1/2}(1, 1, \dots, 1)'$.
	We represent the two samples as the first and second part of a vector $X$, without loss of generality: 
	\begin{align*}
		X
			= 
			\begin{bmatrix}
				X_1 \\
				X_2
			\end{bmatrix}
			=
			\begin{bmatrix}
				1_{m_1}\mu_1 \\
				1_{m_2}\mu_2
			\end{bmatrix}
			+ \widetilde{\epsi},
	\end{align*}
	where $\widetilde{\epsi}$ is an exchangeable random vector, that is, $\widetilde{\epsi} \eqd P\widetilde{\epsi}$ for every permutation matrix $P$.
	We want to test the null hypothesis $H_0 : \mu_1 = \mu_2$ against $H_1 : \mu_1 > \mu_2$.
	
	We now show how this fits into the location model described in Section \ref{sec:location-model}.
	Define $\mu = \tfrac{1}{2} (\mu_1 - \mu_2)$, and let $\iota = (1_{m_1}, -1_{m_2})$ for the remainder of this section.
	Notice that the hypotheses are equivalent to $H_0 : \mu = 0$ and $H_1 : \mu > 0$.
	Define $\epsi = \widetilde{\epsi} + 1_n\tfrac{1}{2} (\mu_1 + \mu_2)$, so that
	\begin{align*}
		X
			&= 
			\begin{bmatrix}
				1_{m_1}\mu_1 \\
				1_{m_2}\mu_2
			\end{bmatrix}
			- 1_n\tfrac{1}{2} (\mu_1 + \mu_2)
			+ \vepsi
			= \viota\mu + \vepsi.
	\end{align*}
	Finally, observe that $\vepsi \eqd P\vepsi$, for every permutation matrix $P$, as addition of a vector with equal elements does not affect exchangeability.

	\subsection{Oracle subgroups for two-sample comparison}	
		In this section, we describe how oracle subgroups of $\calP$ with respect to $\iota$ relate to oracle subgroups of $\calR$ with respect to $1_n$, as described in Section \ref{sec:sign-oracle}.
		In particular, we show how an oracle subgroup of $\calR$ with respect to $1_n$ can be used to construct the matrix representation of an oracle subgroup of $\calP$ with respect to $\iota$, as captured by Theorem \ref{thm:oracle_permutation}.
		This is useful because it is sufficient to have the matrix representation in order to perform the associated group-invariance test.
		Furthermore, such oracle subgroups of $\calR$ with respect to $1_n$ can be found in the database of \citet{Koning_Database_of_Near_2022} and constructed using \citet{Koning_Near_Oracle_Subgroups_2022}.

		\begin{thm}\label{thm:oracle_permutation}
			Let $\calS$ be an oracle subgroup of $\calR$ with respect to $1_n$ and assume that there exists some $R^* \in \calS$ such that $R^*1_n = \viota$.
			Let $\calS^*$ be a subgroup of $\calR$ with elements in $\calS\setminus\{R^*\}$, such that its order is maximal.
			Then, the matrix representation $\fS^*$ of $\calS^*$ is also the matrix representation of an oracle subgroup of $\calP$ with respect to $\viota$.
		\end{thm}
		\begin{proof}
			Observe that $R^*RR^* \in \calS$ for each $R \in \calS$, due to group closure under composition.
			Hence, $\iota'R\iota = 1_n'R^*RR^*1_n = 0$ for all $R \in \calS \setminus\{I\}$, so that $\calS$ is also an oracle subgroup with respect to $\iota$.
			
			Observe that the diagonal elements of $R \in \calS\setminus\{I\}$ are all permutations of each other.
			Taking $\iota = R^*1_n$ as a reference point, multiplication of $\iota$ with any $R \in \calS^*$ coincides with permutation of its elements.
			Notice that $\calS^*$ acts as a group of permutations on the elements of $\iota$.
			This means there exists a subgroup $\widehat{\calS}$ of $\calP$ that is isomorphic to $\calS^*$, such that for each $R \in \calS^*$ there exists a unique $P \in \widehat{\calS}$ such that $R\iota = P\iota$.
			As a consequence, $\iota'P\iota = 0$, for all $P \in \widehat{\calS}\setminus\{I\}$, so that $\widehat{\calS}$ is an oracle subgroup of $\calP$ with respect to $\iota$.
			In addition, the matrix representation $\mathfrak{S}^*$ of $\calS^*$ coincides with the matrix representation of $\widehat{\calS}$.
		\end{proof}
		
		The inclusion of $R^*$ in $\calS$ is without loss of generality in terms of the applicability, as the elements of $\iota$ (and $X$) can simply be rearranged to ensure that $R^* \in \calS$, since each element of $\calS \setminus \{I\}$ has an equal amount of 1s and -1s on the diagonal.
		In addition, note that $2|\calS^*| = |\calS|$ by Proposition \ref{prp:algorithm}.
		
		Finally, we include an example of an oracle subgroup of $\calP$ with respect to $\iota$.
		\begin{exm}
			An example of an oracle subgroup of $\calP$ with respect to $\iota =(1_{n/2}, -1_{n/2})$ for $n = 2m_1 = 4$, is $\calS = \{I, P_1\}$, where 
			\begin{align*}
				P_1 = 
				\begin{bmatrix}
					1 & 0 & 0 & 0 \\
					0 & 0 & 1 & 0 \\
					0 & 1 & 0 & 0 \\
					0 & 0 & 0 & 1
				\end{bmatrix}.
			\end{align*}
			Notice that $P_1$ swaps exactly half of the elements from each sample to the other sample.
			This pattern is a necessary condition for $\calS$ to be an oracle subgroup that generalizes to larger $n$.
			This means we require that $m_1 = m_2$ and $m_1$ is even for an oracle subgroup to exist.
		\end{exm}
		\section{Sign-flipping subgroup algorithm and database}\label{appn:alg}
	In this section, we present an algorithm for construction subgroups $\calS$ of $\calR$ for which $\delta_{\calS}$ or $\delta_{\calS}^{\text{abs}}$ is `small'.
	The main idea behind this algorithm relies on Proposition \ref{prp:algorithm}.
	The result can be summarized as follows: subgroups of $\calR$ are easy to expand to a larger subgroup, and any expansion at least doubles the order of the subgroup.
	
	\begin{prp}\label{prp:algorithm}
		Let $\calA$ be a subgroup of $\calR$ and $R$ an element in $\calR$.
		Then $\calB = \calA \cup R\calA$ is a subgroup of $\calR$, where $R\calA := \{RA\ |\ A \in \calA\}$.
		In addition, if $R \not\in \calA$, then $|\calB| = 2|\calA|$.
	\end{prp}
	
	Based on this result, we propose a simple `greedy' algorithm that constructs near-oracle subgroups of $\calR$.
	This algorithm is presented in Algorithm \ref{alg:near-oracle}.
	The idea of the algorithm is to iteratively expand a subgroup $\calS$ of $\calR$.
	It starts by setting $\calS$ equal to some `good' initial subgroup $\calS^{\text{init}}$ of $\calR$, such as an oracle subgroup that can be found using the constructive proof of Theorem \ref{thm:2primefact}.
	It then considers the expanded subgroup of the form $\calS \cup R\calS$ of $\calR$ for each $R \in \calR \setminus \calS$, where $R\calS := \{RS\ |\ S \in \calS\}$, which is indeed a subgroup by the first part of Proposition \ref{prp:algorithm}.
	Next, it updates the current subgroup $\calS$ to an expanded subgroup that minimizes $\delta_{\calS}$ among the candidate subgroups.
	Note that this minimum is not necessarily unique.
	Finally, the algorithm terminates when the current subgroup $\calS$ is of the desired order. 
	If $M \leq |\calR|$, this algorithm is guaranteed to terminate by the second part of Proposition \ref{prp:algorithm}, as it will eventually expand to the entire group $\calR$.
	
	While the algorithm is based on minimizing $\delta_{\calS}$, there is no guarantee that it terminates at subgroup for which $\delta_{\calS}$ is minimal, but it does guarantee a method of constructing a subgroup of the desired order.
	In addition, the algorithm is not optimal in terms of time complexion, as many of the expansions may be duplicates of each other, but it suffices for our purposes.
	We leave the improvement of the algorithm for future work.
	The performance of group-invariance tests based on subgroups that were found using this algorithm is assessed in Section \ref{sec:simulations}.
	
	Algorithm \ref{alg:near-oracle} has been implemented in the R-package NOS (Near-Oracle Subgroups) \citep{Koning_Near_Oracle_Subgroups_2022}.
	Furthermore, the R-package NOSdata \citep{Koning_Database_of_Near_2022} contains a database of approximately 2500 subgroups for $n \in \{1, \dots, 256\}$ and order $2^{\{0, \dots, 10\}}$.
	For the construction of these subgroups, we drew 100\,000 times without replacement (whenever possible) from $\calR \setminus \calS$ in line 4 of the algorithm. 
	Next to minimizing using $\delta_{\calS}$, it also contains subgroups minimized using $\delta_{\calS}^{\text{abs}}$ for use in two-sided tests.
	
	\begin{algorithm}
		\caption{Near-oracle subgroups}
		\label{alg:near-oracle}
		\begin{algorithmic}[1]
			\State $\calS$ $\gets$ $\calS^{\text{init}}$, \text{ where $\calS^{\text{init}}$ is a subgroup of $\calR$}
			\State $\calC \gets$ $\emptyset$
			\While {$|\calS| < M$}
			\For {$R \in \calR \setminus \calS$}
			\State $\calS^* \gets \calS \cup R\calS$
			\State $\calC \gets \calC \cup \{\calS^*\}$
			\EndFor
			\State $\calS \gets$ $\argmin_{\calS ^* \in \calC} \delta_{\calS^*}$
			\EndWhile
			\State \Return $\calS$
		\end{algorithmic}
	\end{algorithm}
	\newpage
		\section{Proofs}\label{appn:proofs}
	\subsection{Equivalent characterizations of invariance}\label{appn:invariance}
	We say that $X$ is invariant under $\calG$, or $\calG$-invariant, if any of the conditions in the following lemma hold.
	\begin{lem}\label{lem:invariance}
		The following three statements are equivalent.
			\begin{enumerate}
				\item $X \eqd GX$, for all $G \in \calG$,
				\item $X \eqd \overline{G}Y$, for $\overline{G}$ uniform on $\calG$, and some random variable $Y$ on $\calX$ independent of $\overline{G}$,
				\item $X \eqd \overline{G}X$, for $\overline{G}$ uniform on $\calG$, independent of $X$.
			\end{enumerate}
	\end{lem}

	\begin{proof}
		We start with the (1) $\implies$ (3) claim.
		As $X \eqd GX$ for all $G \in \calG$, we have $X \eqd \widetilde{G}X$, where $\widetilde{G}$ can have any distribution on $\calG$.
		Choosing $\widetilde{G}$ to be uniformly distributed on $\calG$ yields the claim.
		Choosing, in addition, $Y$ to be an independent copy of $X$ yields the (1) $\implies$ (2) claim.
		
		For the (2) $\implies$ (1) claim, substituting the equation in (2) into both sides of (1) yields $\overline{G}Y \eqd G\overline{G}Y$, for all $G \in \calG$.
		The claim follows from noticing that $G\overline{G} \eqd \overline{G}$, for any $G \in \calG$.
		
		Furthermore, (3) $\implies$ (2) since we can choose $Y = X$.
	\end{proof}
	\subsection{Proof of Theorem \ref{thm:size_general}}
	First, we prove a lemma, from which the result then follows almost immediately.
	The lemma is presented separately, as it is also used in the proof of Theorem \ref{thm:subgroup_vs_full_group}.
	\begin{lem}\label{lem:size_conditional}
		For every $X$, we have $\SE_{\overline{G}} \phi_{\alpha}^{\calG}(\overline{G}X) \leq \alpha$.
	\end{lem}
	\begin{proof}
		To see why this is true, it is useful to write $\phi_{\alpha}^{\calG}$ in a critical-value threshold form as
		\begin{align*}
			\mathbb{I}\{T(\overline{G}X) > q_X^{\alpha}(\calG)\},
		\end{align*}
		where $q_X^{\alpha}(\calG) := \inf\{z \in \mathbb{R}\ | \SP_{\overline{G}}(T(\overline{G}X) > z) \leq \alpha)\}$ is the $\alpha$-upper quantile of the distribution of $T(\overline{G}X)$ for fixed $X$.
		Then,
		\begin{align*}
			\SE_{\overline{G}}\phi_{\alpha}^{\calG}(\overline{G}X) 
				= \SE_{\overline{G}}\mathbb{I}\{T(\overline{G}X) > q_X^{\alpha}(\calG)\}
				= \SP_{\overline{G}}(T(\overline{G}X) > q_X^{\alpha}(\calG))
				\leq \alpha,
		\end{align*}
		by definition of $q_X^{\alpha}(\calG)$.
	\end{proof}
	
	\begin{proof}[Proof of Theorem \ref{thm:size_general}]
		By Lemma \ref{lem:invariance}, if $X$ is $\calG$-invariant then $X \eqd \overline{G}X$, where $\overline{G}$ is uniform on $\calG$, independently of $X$.
		Then, by Tonelli's theorem and Lemma \ref{lem:size_conditional}, we have
		\begin{align*}
			\SE_X\phi_{\alpha}^{\calG}(X) 
				= \SE_{\overline{G} X}\phi_{\alpha}^{\calG}(\overline{G}X) 
				= \SE_X\SE_{\overline{G}}\phi_{\alpha}^{\calG}(\overline{G}X)
				\leq \SE_{X}\alpha
				= \alpha.
		\end{align*}
	\end{proof}
	
	\subsection{Proof of Theorem \ref{thm:subgroup_vs_full_group}}
	We first prove a lemma, before we proceed with the proof of the theorem.
	
	\begin{lem}\label{lem:subgroup_size_conditional}
		For every $X$, we have $\SE_{\overline{G}}\phi_{\alpha}^{\calS}(\overline{G}X) \leq \alpha$.	
	\end{lem}
	\begin{proof}
		Notice that $\overline{G}$ is $\calG$-invariant.
		From Theorem \ref{thm:subgroup}, we know that this implies $\overline{G}$ is also $\calS$-invariant.
		By Lemma \ref{lem:invariance}, we then have $\overline{G} \eqd \overline{S}\overline{G}$, where $\overline{S}$ and $\overline{G}$ are mutually independent.
		As a consequence,
		\begin{align*}
			\SE_{\overline{G}}\phi_{\alpha}^{\calS}(\overline{G}X)
				= \SE_{\overline{S}\overline{G}}\phi_{\alpha}^{\calS}(\overline{S}\overline{G}X)
				= \SE_{\overline{G}}\SE_{\overline{S}}\phi_{\alpha}^{\calS}(\overline{S}\overline{G}X)
				\leq \SE_{\overline{G}}\alpha
				= \alpha,
		\end{align*}
		where the inequality follows from substituting `$\overline{G}X$' in for `$X$', as well as `$\overline{S}$' for `$\overline{G}$', and `$\calS$' for `$\calG$', in Lemma \ref{lem:size_conditional}.
	\end{proof}

	\begin{proof}[Proof of Theorem \ref{thm:subgroup_vs_full_group}]
		A (sub)group-invariance test can be written in a critical-value threshold form as
		\begin{align*}
			\phi_{\alpha}^{\calS}(X)
				&= \sI\{T(X) > q_X(\calS) \},
		\end{align*}
		where $q_X(\calS) := \inf\{q \in \mathbb{R}\ | \SP_{\overline{S}}(T(\overline{S}X) > q) \leq \alpha)\}$.
		By Lemma \ref{lem:subgroup_size_conditional}, we have that for any $X$ and $\calS$,
		\begin{align*}
			\SE_{\overline{G}}\phi_{\alpha}^{\calS}(\overline{G}X)
				= \SP_{\overline{G}}(T(\overline{G}X) > q_X(\calS))
				\leq \alpha.
		\end{align*}
		Hence, $q_X(\calS) \in \{q \in \mathbb{R}\ | \SP_{\overline{G}}(T(\overline{G}X) > q) \leq \alpha\}$.
		By definition of the infimum, $q_X(\calG) \leq q_X(\calS)$.
		As a consequence, $\phi_{\alpha}^{\calG} \geq \phi_{\alpha}^{\calS}$.
	\end{proof}

	\subsection{Proof of Theorem \ref{thm:root2-tech}}
	\begin{proof}
		Let $\overline{G}$ be uniform on $\calG$ and notice that $X \eqd \iota \mu + \overline{G}\epsi$, by Lemma \ref{lem:invariance}.
		If $\alpha \geq 1 / |\calS|$, then $\phi_{\alpha}^{\calS} \geq \phi_{1/|\calS|}^{\calS}$.
		Then, notice that
		\begin{align*}
			\SE_{\overline{G}} \phi_{1/|\calS|}^{\calS}(\iota \mu + \overline{G}\epsi)
				&= \SP_{\overline{G}}\left(\iota'X > \max_{S \in \calS \setminus\{I\}} \iota'SX\right) \\
				&= \SP_{\overline{G}}\left(\min_{S \in \calS \setminus\{I\} }\iota'X - \iota'SX > 0\right) \\
				&= \SP_{\overline{G}}\left(\min_{S \in \calS \setminus\{I\} } \mu(1 - \iota'S\iota) +(\iota - S\iota)'\overline{G}\epsi > 0\right)
		\end{align*}
		It remains to show that this probability equals 1.
		To show this, it is sufficient to show that the support of $\min_{S \in \calS \setminus\{I\} } \mu(1 - \iota'S\iota) +(\iota - S\iota)'\overline{G}\epsi$ conditional on $\epsi$ is positive.
		
		In particular, it is sufficient (but potentially not necessary) that
		\begin{align}
			\inf_{G \in \calG} \min_{S \in \calS \setminus\{I\} } \mu(1 - \iota'S\iota) +(\iota - S\iota)'G\epsi  > 0.\label{ineq:support}
		\end{align}
		Next, notice that
		\begin{align}
			\inf_{G \in \calG} \min_{S \in \calS \setminus\{I\} }\iota'(I - S)(\iota\mu + \overline{G}\epsi) 
				&= \min_{S \in \calS \setminus\{I\}} \mu(1 - \iota'S\iota) + \inf_{G \in \calG} (\iota - S\iota)'G\epsi \nonumber\\
				&\geq \min_{S \in \calS \setminus\{I\}} \mu(1 - \iota'S\iota) + \inf_{H \in \calH} (\iota - S\iota)'H\epsi \label{ineq:H}\\
				&= \min_{S \in \calS \setminus\{I\}} \mu(1 - \iota'S\iota) - \sqrt{2}\sqrt{1 - \iota'S\iota}\|\epsi\|_2,\nonumber
		\end{align}
		where the inequality follows from the fact that $\calG \subseteq \calH$, and the final equality from showing that the Cauchy-Schwarz inequality is attained by choosing $H$ such that $(\iota - S\iota)'H = \sqrt{2}\sqrt{1 + \iota'S\iota}\epsi$.
		
		The fact that and $\mu, \|\epsi\|_2 \geq 0$ and $\mu \sqrt{1 - \delta_{\calS}} > \sqrt{2}\|\epsi\|_2$ jointly implies that $\sqrt{1 - \iota'S\iota} > 0$ for all $S \in \calS\setminus\{I\}$.
		As a consequence, $\min_{S \in \calS \setminus\{I\}} \mu(1 - \iota'S\iota) + \sqrt{2}\sqrt{1 - \iota'S\iota}\|\epsi\|_2 > 0$ if and only if $\min_{S \in \calS\setminus\{I\}} \mu \sqrt{1 - \iota'S\iota}\mu > \sqrt{2}\|\epsi\|_2$.
		By assumption we have $\mu \sqrt{1 - \delta_{\calS}} > \sqrt{2}\|\epsi\|_2$, this inequality holds if and only if $\min_{S \in \calS\setminus\{I\}} \mu \sqrt{1 - \iota'S\iota}\mu > \sqrt{2}\|\epsi\|_2$.
		This proves the first claim.
		
		For the second claim, notice that as $\alpha = 1/M$ we have $\phi_{\alpha}^{\calS} = \phi_{1/|\calS|}^{\calS}$.
		In addition, as we assume $\calG = \calH$ for the second claim, inequality \eqref{ineq:H} is sharp.
		The only remaining potential loss of sharpness is that inequality \eqref{ineq:support} may not be necessary, which requires some effort to resolve.
		In particular, we will show that as $\calG = \calH$ it can be replaced by a weak inequality, and then we resolve any remaining issues the weak inequality causes.
		
		First, notice that conditional on $\epsi$, $(\iota - S\iota)'\overline{H}\epsi$ has a certain scaled Beta distribution by Lemma \ref{lem:beta}, so that it is a continuous random variable on the real line.
		This means the random variable $\min_{S \in \calS \setminus\{I\}} \mu(1 - \iota'S\iota) + (\iota - S\iota)'\overline{H}\epsi$ is also continuous.
		As a consequence, it does not equal its infimum with $\SP_{\overline{H}}$-probability 1, so that 
		\begin{align*}
			\SP_{\overline{G}}\left(\min_{S \in \calS \setminus\{I\} } \mu(1 - \iota'S\iota) +(\iota - S\iota)'\overline{G}\epsi > 0\right)
				= \SP_{\overline{G}}\left(\min_{S \in \calS \setminus\{I\} } \mu(1 - \iota'S\iota) +(\iota - S\iota)'\overline{G}\epsi \geq 0\right).
		\end{align*}
		This probability equals 1 if and only if $\inf_{G \in \calG} \min_{S \in \calS \setminus\{I\} } \mu(1 - \iota'S\iota) +(\iota - S\iota)'G\epsi \geq 0$.
		Finally, as $\delta_{\calS} < 1$, we have that $\sqrt{1 - \iota'S\iota} > 0$ for all $S\in \calS\setminus\{I\}$.
		This implies $\min_{S \in \calS \setminus\{I\}} \mu(1 - \iota'S\iota) + \sqrt{2}\sqrt{1 - \iota'S\iota}\|\epsi\|_2 \geq 0$ if and only if $\min_{S \in \calS\setminus\{I\}} \mu \sqrt{1 - \iota'S\iota}\mu \geq \sqrt{2}\|\epsi\|_2$, which proves the result.
	\end{proof}

	\begin{rmk}\label{rmk:root2-discussion}
		Theorem \ref{thm:root2-tech} may seem unusual, as the claims are conditional on $\|\epsi\|_2$ and because they consider non-asymptotic consistency, as opposed to asymptotic consistency.
		However, these components are natural to a group-invariance setting, as inference is based only on the invariance of $\epsi$: the fact that $\epsi \eqd G\epsi$, for all $G \in \calG$.
		
		In particular, the map $\epsi \mapsto \|\epsi\|_2$ is invariant under $\calH$, as $\|\epsi\|_2 = \|H\epsi\|_2$, for any $H \in \calH$.
		In fact, it is the so-called maximal invariant of $\calH$.
		Intuitively speaking, this means it contains all the information in $\epsi$ that cannot be exploited by inference based on any subgroup of $\calH$.
		As a consequence, we should expect any result based on invariance under $\calG$ to hold conditional on $\|\epsi\|_2$.  
		
		To understand the appearance of non-asymptotic consistency, notice that $\calG$ is compact and $T$ is continuous.
		As a consequence, conditional on $X$, the support of the random variable $T(\overline{G}X)$ is compact, even if its unconditional support is not.
		If the entire support of a compactly supported distribution exceeds a threshold, the random variable exceeds the threshold with probability one.
		The proof strategy relies on reducing the power of the test to such a threshold condition.
	\end{rmk}
		
	\begin{rmk}\label{rmk:sharpness}
		An inspection of the proof of Theorem \ref{thm:root2-tech} shows that if $\alpha = 1 / |\calS|$ and $\calG \neq \calH$, then the only step with a serious loss of sharpness seems to be inequality \eqref{ineq:H}, which in essence comes down to the Cauchy-Schwarz inequality.
		
		The main difficulty in obtaining a sharper (non-trivial) statement for arbitrary groups $\calG \neq \calH$ lies in the fact that the term $\inf_{G \in \calG} (\iota - S\iota)'G\epsi$ will no longer rely on $S$ just through $\iota'S\iota$ and on $\epsi$ through $\|\epsi\|_2$, but also through the interaction between $S$ and $\epsi$.
		In particular, assuming $\|\epsi\|_2 > 0$,
		\begin{align*}
			\inf_{G \in \calG} (\iota - S\iota)'G\epsi
				= \sqrt{2}\sqrt{1 - \iota'S\iota}\|\epsi\|_2 \inf_{G \in \calG}\frac{(\iota - S\iota)'G\epsi}{\sqrt{2}\sqrt{1 - \iota'S\iota}\|\epsi\|_2},
		\end{align*}		
		where the final term involving the infimum typically does not reduce to $-1$.
		As a consequence, the first claim in Theorem \ref{thm:root2-tech} can be sharpened to
		\begin{itemize}
			\item If $\alpha \geq 1 / |\calS|$ and $\min_{S \in \calS\setminus\{I\}} \frac{\mu}{\|\epsi\|_2}(1 - \iota'S\iota) + \sqrt{2}\sqrt{1 - \iota'S\iota}\inf_{G \in \calG}\frac{(\iota - S\iota)'G\epsi}{\sqrt{2}\sqrt{1 - \iota'S\iota}\|\epsi\|_2} > 0$, then $\SE_{\overline{G}} \phi_{\alpha}^{\calS}(X) = 1$.	
		\end{itemize}
		Unfortunately, this term no longer explicitly depends on the leak, so that the impact of the leak on the power is hard to ascertain.
		For this reason, we do not include the sharper statement in the main text.
		However, a sharper and clean statement may perhaps be obtained in specific applications, if one can find a lower bound for $\inf_{G \in \calG}\frac{(\iota - S\iota)'G\epsi}{\sqrt{2}\sqrt{1 - \iota'S\iota}\|\epsi\|_2}$.
	\end{rmk}
	\subsection{Proof of Theorem \ref{thm:root2-tech-abs}}
	The proof of Theorem \ref{thm:root2-tech-abs} is similar to the proof of Theorem \ref{thm:root2-tech}, but requires some extra work to deal with the absolute values.
	
	\begin{proof}
		As in the proof of Theorem \ref{thm:root2-tech}, let $\overline{G}$ be uniform on $\calG$ and so that $X \eqd \iota \mu + \overline{G}\epsi$, by Lemma \ref{lem:invariance}.
		
		Our strategy will be to first get rid of the absolute value signs, which will then allow us to apply the same steps as used in the proof in Theorem \ref{thm:root2-tech}.
		First, define $\calS^+ = \calS \cup \{-S\ |\ S\in \calS\}$, which is still a subset of $\calH$.
		Using this notation, we have that
		\begin{align*}
			\max_{S \in \calS\setminus\{I\}} |\iota'SX| = \max_{S \in \calS^+\setminus\{I\}} \iota'SX.
		\end{align*}
		
		Second, observe that $\frac{|\mu|}{\|\epsi\|_2} \geq \frac{|\mu|}{\|\epsi\|_2}\sqrt{1 - \delta_{\calS}^{\text{abs}}} > \sqrt{2}$ since $\delta_{\calS}^{\text{abs}} \leq 1$.
		This allows us to consider two cases: $\frac{\mu}{\|\epsi\|_2} > \sqrt{2}$ and $\frac{\mu}{\|\epsi\|_2} < -\sqrt{2}$.
		We start with the first case, which implies
		\begin{align*}
			 \iota'X = \mu + \iota'\overline{G}\epsi \geq \mu + \inf_{G \in \calH} \iota'G\epsi = \mu + \|\epsi\|_2 > 0,
		\end{align*}
		so that $|\iota'X| = \iota'X$.
		
		Using the steps analogous to those in the proof of Theorem \ref{thm:root2-tech}, yields the sufficient condition $\mu\sqrt{1 - \delta_{\calS^+}} > \sqrt{2}\|\epsi\|_2$.
		Similarly, the case that $\frac{\mu}{\|\epsi\|_2} < -\sqrt{2}$ yields $-\mu\sqrt{1 - \delta_{\calS^+}} > \sqrt{2}\|\epsi\|_2$.
		Combining these inequalities and observing that $\delta_{\calS^+} = \delta_{\calS}^{\text{abs}}$ yields the condition $|\mu|\sqrt{1 - \delta_{\calS}^{\text{abs}}} > \sqrt{2}\|\epsi\|_2$.
		This proves the first claim.
		
		For the second claim, we can follow the same steps as in the proof of Theorem \ref{thm:root2-tech}, after observing that Lemma \ref{lem:atrocity} implies that we can analyze the $\iota'X > 0$ and $\iota'X < 0$ cases separately when we assume $\SE_{\overline{H}} \overline{\psi}_{1/|\calS|}^{\calS}(\iota \mu + \overline{H}\epsi) = 1$.
		Here, the $\iota'X = 0$ case can be ignored, since $\iota'X$ is a continuous random variable. \\
	\end{proof}
	
	The proof of the second claim in Theorem \ref{thm:root2-tech-abs} relies on the following lemma, for which we unfortunately only have a rather tedious proof.
	\begin{lem}\label{lem:atrocity}
		If
		\begin{align*}
			\SP_{\overline{H}}\left(\min_{S \in \calS \setminus\{I\} } |\iota'X| - |\iota'SX| > 0\right) = 1,
		\end{align*}
		then either
		\begin{align*}
			\SP_{\overline{H}}\left(\min_{S \in \calS \setminus\{I\} } \iota'X - |\iota'SX| > 0\right) = 1,
		\end{align*}
		or
		\begin{align*}
			\SP_{\overline{H}}\left(\min_{S \in \calS \setminus\{I\} } -\iota'X - |\iota'SX| > 0\right) = 1.
		\end{align*}
	\end{lem}
	\begin{proof}			
		To prove the claim, we prove the contrapositive.
		In particular, we assume that both
		\begin{align*}
			\SP_{\overline{H}}\left(\min_{S \in \calS \setminus\{I\}}\min_{\sigma \in \{-1, 1\}} \iota'X - \sigma\iota'SX > 0\right) > 0, \\
			\SP_{\overline{H}}\left(\min_{S \in \calS \setminus\{I\}}\min_{\sigma \in \{-1, 1\}} -\iota'X - \sigma\iota'SX > 0\right) > 0,
		\end{align*}
		and show that this implies $\mu/\|\epsi\|_2 \in (-1, 1)$.
		To do so, we write the inequalities as
		\begin{align*}
			\SP_{\overline{H}}\left(\min_{S \in \calS \setminus\{I\}}\min_{\sigma \in \{-1, 1\}} \mu(1 - \sigma\iota'S\iota) + (\iota - \sigma S\iota)'\overline{H}\epsi > 0\right) > 0, \\
			\SP_{\overline{H}}\left(\min_{S \in \calS \setminus\{I\}}\min_{\sigma \in \{-1, 1\}} \mu(-1 - \sigma\iota'S\iota) + (-\iota - \sigma S\iota)'\overline{H}\epsi > 0\right) > 0.
		\end{align*}
		Notice that this implies
		\begin{align*}
			\sup_{H \in \calH} \min_{S \in \calS \setminus\{I\}}\min_{\sigma \in \{-1, 1\}} \mu(1 - \sigma\iota'S\iota) + (\iota - \sigma S\iota)'H\epsi > 0, \\
			\sup_{H \in \calH} \min_{S \in \calS \setminus\{I\}}\min_{\sigma \in \{-1, 1\}} \mu(-1 - \sigma\iota'S\iota) + (-\iota - \sigma S\iota)'H\epsi> 0.
		\end{align*}
		By the max-min inequality, this implies
		\begin{align*}
			\min_{S \in \calS \setminus\{I\}}\min_{\sigma \in \{-1, 1\}} \mu(1 - \sigma\iota'S\iota) + \sup_{H \in \calH} (\iota - \sigma S\iota)'H\epsi > 0, \\
			\min_{S \in \calS \setminus\{I\}}\min_{\sigma \in \{-1, 1\}} \mu(-1 - \sigma\iota'S\iota) + \sup_{H \in \calH} (-\iota - \sigma S\iota)'H\epsi > 0.
		\end{align*}
		Using the Cauchy-Schwarz inequality, this is equivalent to
		\begin{align*}
			\min_{S \in \calS \setminus\{I\}}\min_{\sigma \in \{-1, 1\}} \mu(1 - \sigma\iota'S\iota) + \sqrt{2}\sqrt{1 - \sigma\iota'S\iota}\|\epsi\|_2 > 0, \\
			\min_{S \in \calS \setminus\{I\}}\min_{\sigma \in \{-1, 1\}} \mu(-1 - \sigma\iota'S\iota) + \sqrt{2}\sqrt{1 + \sigma\iota'S\iota}\|\epsi\|_2  > 0.
		\end{align*}
		Which is equivalent to
		\begin{align*}
			\min_{S \in \calS \setminus\{I\}}\min_{\sigma \in \{-1, 1\}} \frac{\mu}{\|\epsi\|_2}\sqrt{1 - \sigma\iota'S\iota} > -\sqrt{2}, \\
			\min_{S \in \calS \setminus\{I\}}\min_{\sigma \in \{-1, 1\}} -\frac{\mu}{\|\epsi\|_2}\sqrt{1 + \sigma\iota'S\iota} > -\sqrt{2},
		\end{align*}
		which implies $\mu/\|\epsi\|_2 \in (-1, 1)$, since $-1 \leq \sigma\iota'S\iota \leq 1$.
		
		It remains to show that this implies
		\begin{align*}
			\SP_{\overline{H}}\left(\min_{S \in \calS \setminus\{I\} } |\iota'X| - |\iota'SX| > 0\right) < 1,
		\end{align*}
		or equivalently
		\begin{align*}
			\SP_{\overline{H}}\left(\min_{S \in \calS \setminus\{I\}}\min_{\sigma \in \{-1, 1\}}\max_{\tau \in \{-1, 1\}} \tau\iota'X - \sigma\iota'SX < 0\right) > 0,
		\end{align*}
		which is implied by 
		\begin{align*}
			\SP_{\overline{H}}\left(\tau\iota'X - \sigma\iota'SX < 0\right) > 0, \forall \sigma, \tau, S.
		\end{align*}
		Notice that 
		\begin{align*}
			\SP_{\overline{H}}\left(\tau\iota'X - \sigma\iota'SX < 0\right)
				&= \SP_{\overline{H}}\left(\mu(\tau - \sigma\iota'S\iota) + (\tau\iota - \sigma S\iota)'\overline{H}\epsi < 0\right) \\
				&= \SP_{\overline{H}}\left(\tfrac{\mu}{\|\epsi\|_2}\tfrac{\tau - \sigma\iota'S\iota}{\sqrt{2}\sqrt{1 - \sigma\iota'S\iota}} < \tfrac{(\tau\iota - \sigma S\iota)'}{\sqrt{2}\sqrt{1 - \sigma\iota'S\iota}}\overline{H}\tfrac{\epsi}{\|\epsi\|_2}\right) \\
				&= 1 - F^{\beta\left(\tfrac{n-1}{2}, \tfrac{n-1}{2}, -1, 1\right)}\left(\tfrac{\mu}{\|\epsi\|_2}\tfrac{\tau - \sigma\iota'S\iota}{\sqrt{2}\sqrt{1 - \sigma\iota'S\iota}}\right) \\
				&> 1 - F^{\beta\left(\tfrac{n-1}{2}, \tfrac{n-1}{2}, -1, 1\right)}(1)
				= 0,
		\end{align*}
		where the third equality follows from Lemma \ref{lem:beta}, and the inequality from the continuity of the beta distribution on the support $[-1, 1]$ and the fact that $-1 < \tfrac{\mu}{\|\epsi\|_2}\tfrac{\tau - \sigma\iota'S\iota}{\sqrt{2}\sqrt{1 - \sigma\iota'S\iota}} < 1$, for all $\sigma, \tau$ and $S \in \calS\setminus\{I\}$.
		This proves the claim.
	\end{proof}

	\subsection{Proof of Theorem \ref{thm:root2-MC}}
	For the proof of Theorem \ref{thm:root2-MC}, we use the following lemma that is similar to Lemma 4.2 in \citet{dobriban2021consistency}.
	
	\begin{lem}\label{lem:max}
		Suppose $M$ is fixed. 
		Let $Z_i$, $i \in \{1, \dots, M\}$ be identically distributed real-valued random variables and let $Z$ be some other real-valued random variable.
		Then
		\begin{align*}
			\SP_{Z_1, \dots Z_M}(Z > \max_{i} Z_i) = 1,
		\end{align*}
		if and only if $\SP_{Z_1}(Z > Z_1) = 1$.
	\end{lem}
	\begin{proof}
		Following Lemma 4.2 of \citet{dobriban2021consistency}, consider the events $A_i = \{Z \leq Z_i\}$.
		Taking complements, it is sufficient to show $\SP_{A_1, \dots A_M}(\bigcup_{i=1}^M A_i) = 0$ if and only if $\SP_{A_1}(A_1) = 0$.
		By the union bound we have
		\begin{align*}
			\SP_{A_1}(A_1)
				\leq \SP_{A_1, \dots A_M}(\bigcup_{i=1}^M A_i)
				\leq \sum_{i = 1}^M \SP_{A_i}(A_i)
				= M \SP_{A_1}(A_1),
		\end{align*}
		where the final equality follows from the fact that the $Z_i$ are identically distributed.
		The claim follows from observing that the inequalities imply $\SP_{A_1, \dots A_M}(\bigcup_{i=1}^M A_i) = 0$ if and only if $\SP_{A_1}(A_1) = 0$.
	\end{proof}
		
	\begin{proof}[Proof of Theorem \ref{thm:root2-MC}]			
		For the second claim, notice that by the first claim in Theorem \ref{thm:root2-tech},
		\begin{align*}
			\sI\left\{\mu\sqrt{1 - \delta_{\calG_M}} \geq \sqrt{2}\|\epsi\|_2\right\}
				\leq \sI\left\{\SE_{\overline{G}}\phi_{\alpha}^{\calG_M}(X) = 1\right\}.
		\end{align*}
		Taking the expectation with respect to $\calG_M$ on both sides yields
		\begin{align*}
			\SP_{\calG_M}(\mu \sqrt{1 - \delta_{\calG_M}} \geq \sqrt{2}\|\epsi\|_2)
				\leq \SP_{\calG_M}(\SE_{\overline{G}}\phi_{\alpha}^{\calG_M}(X) = 1).
		\end{align*}
		As $\SE_{\overline{G}}\phi_{\alpha}^{\calG_M}(X) \leq 1$, the right-hand side of this display equals 1 if and only if
		\begin{align*}
			\SE_{\calG_M}\SE_{\overline{G}}\phi_{\alpha}^{\calG_M}(X) = 1.
		\end{align*}
		The left-hand side is equal to 
		\begin{align*}
			\SP_{\calG_M}(\mu \sqrt{1 - \delta_{\calG_M}} \geq \sqrt{2}\|\epsi\|_2)
				&= \SP_{\calG_M}(1 - \tfrac{\|\epsi\|_2^2} {\mu^2}\geq \max_{G \in \calG_M \setminus \{I\}} \iota'G\iota)
		\end{align*}
		as $\mu \geq 0$.
		By Lemma \ref{lem:max}, the final expression in the previous display equals 1 if and only if 
		\begin{align*}
			\SP_{\overline{G}}(\mu^2 - \|\epsi\|_2^2\geq \mu^2\iota'\overline{G}\iota) = 1.
		\end{align*}
		Which is true if and only if 
		\begin{align*}
			\mu\sqrt{1 - \delta_{\calG}} \geq \sqrt{2}\|\epsi\|_2.
		\end{align*}
		This proves the first claim.
		The second claim follows from applying the same reasoning to the second claim of Theorem \ref{thm:root2-tech}.
	\end{proof}
	
	\subsection{A power conjecture}\label{sec:conj}
		Figure \ref{fig:A} shows that MC group-invariance tests can outperform oracle subgroup-invariance tests.
		However, beyond the $n = 2$ case, we have not been able to find any counterexamples.
		An in-depth investigation of the $n = 2$ case showed that this phenomenon relies on the fact that small spherical caps contain a large amount of the mass of the sphere, which can be seen in the fact that the density of the $\text{Beta}(\tfrac{n - 1}{2}, \tfrac{n-1}{2}, -1, 1)$-distribution is U-shaped.
		This U-shape is no longer present if $n \geq 3$.
		Unfortunately, we were not able to construct a proof.
		
		This leads us to
			
		\begin{cnj}\label{cnj:n>=3}
			Suppose that $\epsi$ is $\calH$-invariant $\calS$ is an oracle subgroup of $\calH$ of order $M$, $n \geq 3$ and let $\|\epsi\|_2 > 0$ be fixed.
			Then, $\SE_{\overline{H}}\phi_\alpha^{\calS}(\iota\mu + \overline{H}\epsi) \geq \SE_{\calH_M}\SE_{\overline{H}}\phi_\alpha^{\calH_M}(\iota\mu + \overline{H}\epsi)$.
		\end{cnj}
		
		\begin{rmk}
			If Conjecture \ref{cnj:n>=3} is true for every $\|\epsi\|_2$, it immediately follows that it is true for random $\|\epsi\|_2$ as well. 
		\end{rmk}
	
	\subsection{Proof of Theorem \ref{thm:t-test}}
	In order to present the proof of Theorem \ref{thm:t-test}, we first prove two simple lemmas.
	Lemma \ref{lem:beta} is also used in several other proofs.
		
	\begin{lem}\label{lem:beta}
		Let $a$ and $b$ be unit $n$-vectors and let $\overline{H}$ be uniform on $\calH$.
		Then, $a'\overline{H}b \sim \textnormal{Beta}(\frac{n-1}{2}, \frac{n-1}{2}, -1, 1)$.
	\end{lem}
	\begin{proof}
		First, notice that by the third characterization in Lemma \ref{lem:invariance}, we have $\overline{H} \eqd \overline{H}_1\overline{H}$, where $\overline{H}_1$ is uniform on $\calH$, independent of $\overline{H}$.
		Choosing $\overline{H}_2$ to uniform on $\calH$ independent of $\overline{H}$ and $\overline{H}_1$, we have $\overline{H} \eqd \overline{H}_1 \overline{H}_2$.
		As a consequence, $a'\overline{H}b \eqd a'\overline{H}_1\overline{H}_2b \eqd \overline{u}'\overline{v}$, where $\overline{u}$ and $\overline{v}$ are independent and uniform random $n$-vectors on the unit sphere.
	
		Suppose $y$ and $z$ are independent multivariate standard normally distributed $n$-vectors.
		The random vectors $u_y := y/\|y\|_2$ and $u_z := z/\|z\|_2$ are $\calH$-invariant unit  $n$-vectors.
		Hence, $\overline{u}'\overline{v} \eqd u_y'u_z$.
		
		Finally, notice that $u_y'u_z$ is the sample correlation coefficient between two independent multivariate standard normally distributed $n$-vectors, which is $\text{Beta}(\frac{n-1}{2}, \frac{n-1}{2}, -1, 1)$-distributed \citep{fisher1915frequency, hotelling1953new}.
		This proves the claim.
	\end{proof}
	
	Define the maps $f_n : (-1, 1) \to \mathbb{R}$ as $y \mapsto \sqrt{n-1} \tfrac{y}{\sqrt{1 - y^2}}$, and notice that $f_n(y)$ is strictly increasing in $y \in \mathbb{R}$, $n \in \mathbb{N}^+$.
			
	\begin{lem}\label{lem:beta_t}
		If $Z \sim \textnormal{Beta}(\tfrac{n-1}{2}, \tfrac{n-1}{2}, -1, 1)$, then $f(Z) \sim t_{n-1}$.	
	\end{lem}
	\begin{proof}
		Suppose $\eta \sim \calN(0, \sigma^2I)$, so that $\eta / \|\eta\|_2$ is uniform on the unit sphere and so $\iota'\eta/\|\eta\|_2 \eqd Z$ by Lemma \ref{lem:beta}.
		Some straightforward algebra shows that
		\begin{align*}
			f(\iota'\eta/\|\eta\|_2) 
				= \sqrt{n - 1} \frac{\iota'\eta}{\sqrt{\eta'(I - \iota\iota')\eta}},
		\end{align*}
		which well-known to have a $t_{n-1}$ distribution.
		Hence, $f(Z) \sim t_{n-1}$.
	\end{proof}

	\begin{rmk}
		Similar results exist in the literature.
		For example \citet{efron1969student} shows that $f(a'\overline{H}b) \sim t_{n-1}$, by noticing that geometric arguments made by \citet{fisher1925applications} for the normal distribution only rely on the orthogonal invariance of the distribution.	
	\end{rmk}

	Combining Lemmas \ref{lem:beta} and \ref{lem:beta_t} yields the following straightforward proof of Theorem \ref{thm:t-test}.
	
	\begin{proof}
		We exclude the case that $X = 0$, where the $t$-test is poorly defined. 
		Let $Z$ denote a $\textnormal{Beta}(\tfrac{n-1}{2}, \tfrac{n-1}{2}, -1, 1)$-distributed random variable, and $\tau$ a $t_{n-1}$ distributed random variable.
		Then,
		\begin{align*}
			\phi_{\alpha}^{\calH}(X)
				&:= \sI\left\{\SP_{\overline{H}}(\iota'\overline{H}X \geq \iota'X) \leq \alpha\right\} \\
				&= \sI\left\{\SP_{\overline{H}}\left(\frac{\iota'\overline{H}X}{\|X\|_2}  \geq \frac{\iota'X}{\|X\|_2}\right) \leq \alpha\right\} \\
				&= \sI\left\{\SP_{Z}\left(Z \geq \frac{\iota'X}{\|X\|_2}\right) \leq \alpha\right\} \\
				&= \sI\left\{\SP_{\tau}\left(\tau \geq f\left(\frac{\iota'X}{\|X\|_2}\right)\right) \leq \alpha\right\} \\
				&= \sI\left\{\SP_{\tau}\left(\tau \geq \sqrt{n-1}\frac{\iota'X}{\sqrt{X'(I - \iota\iota')X}}\right) \leq \alpha\right\} \\
				&=: \sI\left\{\sqrt{n-1}\frac{\iota'X}{\sqrt{X'(I - \iota\iota')X}} > t_{n-1}^{\alpha}\right\},
		\end{align*}
		where the second equality follows from Lemma \ref{lem:beta} and the third equality from Lemma \ref{lem:beta_t} and the fact that $f$ is strictly increasing.
	\end{proof}

	\subsection{Proof of Corollary \ref{cor:MC_t}}
	\begin{proof}
		This proof is analogous to the proof of Theorem \ref{thm:t-test} after noticing that one can use Lemmas \ref{lem:beta} and \ref{lem:beta_t} to transform a sample of independent uniform draws from $\overline{H}$ into a sample of independent draws from the $t_{n-1}$-distribution and use the sample quantile as a critical value.
	\end{proof}
	\subsection{Proof of Theorem \ref{thm:norm}}
	\begin{proof}
		We start by assuming that $M = n$.
		Let $\mathfrak{S}$ be the matrix representation of $\calS$, which exists by Proposition \ref{prp:low_leak}.
		Without loss of generality, let $\iota$ be the first column of $\mathfrak{S}$.
		From Proposition \ref{prp:leak_representation}, we know that all columns of $\mathfrak{S}$ are orthogonal.
		Hence, $\mathfrak{S}'\iota = e_1 = (1, 0, \dots 0)'$. 
		This yields
		\begin{align*}
			\fS'x
				= \mu\mathfrak{S}'\iota + \mathfrak{S}'\epsi
				= \mu e_1  + \mathfrak{S}'\epsi
				\eqd \mu e_1  + \epsi,
		\end{align*}
		where the final step follows from the first statement in Lemma \ref{lem:invariance}, the fact that $\fS'$ is an orthonormal matrix and the $\calH$-invariance of $\epsi$.
		The $\calS$-invariance test compares the first element of the vector $(\mu e_1  + \epsi)$ to its remaining elements.
		As the elements of $\epsi$ are i.i.d., this is exactly a Monte Carlo $Z$-test.
		
		To prove the $M \leq n$ case, one can pad the resulting  $n \times M$ matrix representation with zero-columns until it is $n \times n$ and follow similar steps, noting that the zero-columns wil eliminate some elements of $\epsi$.
	\end{proof}
	
	\begin{rmk}\label{rmk:norm_nec}
		Normality is almost necessary for the proof strategy in Theorem \ref{thm:norm}, for the following reason.
		For the proof, we require that $\fS'\epsi \eqd \epsi$, and $\epsi$ has i.i.d. elements.
		Let $s_1$ and $s_2$ denote two columns of $\fS$.
		Then $(s_1, s_2)'\vepsi \eqd (e_1, e_2)'\vepsi$.
		As $e_1'\epsi$ and $e_2'\epsi$ are independent, so are $s_1'\epsi$ and $s_2'\epsi$.
		
		This brings us very close to the conditions of the Darmois-Skitovich Theorem \citep{darmois1953analyse, skitovitch1953property}.
		This theorem states the following: $s_1$ and $s_2$ have only non-zero elements, $\epsi$ has i.i.d. elements and $s_1'\epsi$ and $s_2'\epsi$ are independent if and only if $\epsi \sim N(0, \sigma^2I)$, for some $\sigma^2 > 0$.
		Therefore, if $\fS$ has two columns with non-zero elements, then we indeed require that $\epsi \sim \calN(0, \sigma^2I)$.
	\end{rmk}

	\subsection{Proof of Proposition \ref{prp:low_leak}}
	\begin{proof}
		Start by noticing that $\sup_{S_1 \neq S_2 \in \calS\setminus\{I\}} \iota'S_1'S_2\iota \leq \delta_{\calS}$, as $S_1S_2 \in \calS$. 
		If $\delta_{\calS} < 1$, then $\calS \mapsto \calS\iota$ is a bijection.
		If it were not a bijection, then we could find two distinct elements $S_1, S_2 \in \calS$ for which $S_1\iota = S_2\iota$.
		However, this would imply $\iota'S_2'S_1\iota = 1$, which contradicts the premise that $\delta_{\calS} < 1$.
		
		Suppose, for the sake of contradiction, that $\calS$ is infinite.
		As $\calS \mapsto \calS\iota$ is a bijection, $\calS\iota$ is also infinite.
		Notice that the elements of $\calS\iota$ are points on the unit sphere.
		As $\sup_{S_1 \neq S_2 \in \calS\setminus\{I\}} \iota'S_1'S_2\iota \leq \delta_{\calS} < 1$, all these points are at least some fixed distance away from each other.
		Hence, no sequence of points in $\calS\iota$ has a convergent subsequence.
		This contradicts the sequential compactness of the unit sphere in $n$ dimensions.
		Hence, $\calS$ is finite.
	\end{proof}
	\subsection{Proof of Proposition \ref{prp:existence}}
	We first prove two lemmas, which we then combine to prove Theorem \ref{prp:existence}.
	The first lemma constructs oracle subgroups of any order $p$ for a specific choice of $\iota$.
	The second lemma transforms these oracle subgroups to oracle subgroups for any choice of $\iota$.

\begin{lem}\label{lem:cyclic}
	For each $p$, $1 \leq p \leq n$, there exists a cyclic subgroup $\calS$ of the permutation group $\calP$ of order $p$ that is an oracle subgroup of $\calH$ with respect to $\ve_1$.
\end{lem}
\begin{proof}
	If $p = 1$, then the group consists only of the identity element, which is indeed an oracle subgroup.
	For $p > 1$, without loss of generality, let
	\begin{align*}
		S_1 = (e_p, e_1, e_2, \dots, e_{p-1}, e_{p+1}, e_{p+2}, \dots, e_n).
	\end{align*}
	The matrix $S_1$ is a generator of a cyclic subgroup $\calS$ of $\calP$ of order $p$.
	It is easily verified that $e_1'Se_1 = 0$ for all $S \in \calS \setminus \{I\}$.
	Hence, $\calS$ is an oracle subgroup of $\calH$ of order $p$ with respect to $e_1$.
\end{proof}
	
\begin{lem}\label{lem:oracle_transform}
	Let $a$ and $b$ be unit vectors.
	Let $Q$ be an orthonormal matrix such that $Qa= b$, which exists.
	Let $\calS$ be an oracle subgroup of $\calH$ with respect to $a$.
	Then, $\calG$ with elements $G = QSQ'$, $S \in \calS$, is an oracle subgroup of $\calH$ with respect to $b$, and $\calG$ is isomorphic to $\calS$.	
\end{lem}
\begin{proof}
	We have that $\calG \subset \calH$, as the elements of $\calG$ are compositions of orthonormal matrices.
	Furthermore, it is a sub\emph{group}, as
	\begin{itemize}
		\item $I \in \calS$, so $QIQ' = I \in \calG$,
		\item for any $G \in \calG$ and some $S \in \calS$, we have $G' = (QSQ')' = QS'Q'  \in \calG$,
		\item for any $G_1,G_2 \in \calG$ and some $S_1, S_2 \in \calS$, we have $G_1\_2 = QS_1Q'QS_2Q' = QS_1S_2Q' \in \calG$.
	\end{itemize}
	In addition, it is an oracle subgroup with respect to $b$ as $b'Gb = b'QSQ'b = a'Sa = 0$, for any $G \in \calG \setminus \{I\}$ and some $S \in \calS$.
	Finally, $\calG$ and $\calS$ are isomorphic as the map $S \mapsto QSQ'$, $S \in \calS$, $G \in \calG$, is a bijection from $\calS$ to $\calG$.
\end{proof}

\begin{proof}[Proof of Theorem \ref{prp:existence}]
	From Lemma \ref{lem:cyclic} we can obtain an oracle subgroup $\calS$ of $\calH$ with respect to $e_1$ of any desired order $p$, $1 \leq p \leq n$.
	Using Lemma \ref{lem:oracle_transform}, we can transform $\calS$ into an isomorphic oracle subgroup of $\calH$ with respect to any unit vector.
\end{proof}
	\end{appendix}
	
	\bibliographystyle{abbrvnat}
	\bibliography{bibliogr.bib}   

\begin{thebibliography}{63}
\providecommand{\natexlab}[1]{#1}
\providecommand{\url}[1]{\texttt{#1}}
\expandafter\ifx\csname urlstyle\endcsname\relax
  \providecommand{\doi}[1]{doi: #1}\else
  \providecommand{\doi}{doi: \begingroup \urlstyle{rm}\Url}\fi

\bibitem[Anderson and Robinson(2001)]{anderson2001permutation}
M.~J. Anderson and J.~Robinson.
\newblock Permutation tests for linear models.
\newblock \emph{Australian \& New Zealand Journal of Statistics}, 43\penalty0
  (1):\penalty0 75--88, 2001.

\bibitem[Andreella(2021)]{Andreella_fMRIdata_2021}
A.~Andreella.
\newblock {fMRIdata}.
\newblock \url{https://github.com/angeella/fMRIdata}, 2021.

\bibitem[Andreella et~al.(2020)Andreella, Hemerik, Weeda, Finos, and
  Goeman]{andreella2020permutation}
A.~Andreella, J.~Hemerik, W.~Weeda, L.~Finos, and J.~Goeman.
\newblock Permutation-based true discovery proportions for fmri cluster
  analysis.
\newblock \emph{arXiv preprint arXiv:2012.00368}, 2020.

\bibitem[Bekker and Lawford(2008)]{bekker2008symmetry}
P.~A. Bekker and S.~Lawford.
\newblock Symmetry-based inference in an instrumental variable setting.
\newblock \emph{Journal of econometrics}, 142\penalty0 (1):\penalty0 28--49,
  2008.

\bibitem[Berry et~al.(2014)Berry, Johnston, and Mielke~Jr]{berry2014chronicle}
K.~J. Berry, J.~E. Johnston, and P.~W. Mielke~Jr.
\newblock A chronicle of permutation statistical methods.
\newblock \emph{Cham: Springer}, 2014.

\bibitem[Blain et~al.(2022)Blain, Thirion, and Neuvial]{blain2022notip}
A.~Blain, B.~Thirion, and P.~Neuvial.
\newblock Notip: Non-parametric true discovery proportion control for brain
  imaging.
\newblock \emph{NeuroImage}, 260:\penalty0 119492, 2022.

\bibitem[Blanchard et~al.(2020)Blanchard, Neuvial, and
  Roquain]{blanchard2020post}
G.~Blanchard, P.~Neuvial, and E.~Roquain.
\newblock Post hoc confidence bounds on false positives using reference
  families.
\newblock \emph{The Annals of Statistics}, 48\penalty0 (3):\penalty0
  1281--1303, 2020.

\bibitem[Chmielewski(1981)]{chmielewski1981elliptically}
M.~Chmielewski.
\newblock Elliptically symmetric distributions: A review and bibliography.
\newblock \emph{International Statistical Review/Revue Internationale de
  Statistique}, pages 67--74, 1981.

\bibitem[Clemens(2021)]{clemens2021enhancing}
J.~Clemens.
\newblock Enhancing the power of permutation tests for positive serial
  dependence in binary data by using streak-breaking subgroups.
\newblock Master's thesis, Dec. 2021.
\newblock URL \url{http://hdl.handle.net/2105/60882}.

\bibitem[Conway and Sloane(1998)]{conway1998sphere}
J.~H. Conway and N.~J.~A. Sloane.
\newblock \emph{Sphere packings, lattices and groups (Third Edition)}.
\newblock Springer-Verlag, New York, 1998.

\bibitem[Darmois(1953)]{darmois1953analyse}
G.~Darmois.
\newblock Analyse g{\'e}n{\'e}rale des liaisons stochastiques: etude
  particuli{\`e}re de l'analyse factorielle lin{\'e}aire.
\newblock \emph{Revue de l'Institut International de Statistique}, pages 2--8,
  1953.

\bibitem[Davidson and Flachaire(2008)]{davidson2008wild}
R.~Davidson and E.~Flachaire.
\newblock The wild bootstrap, tamed at last.
\newblock \emph{Journal of Econometrics}, 146\penalty0 (1):\penalty0 162--169,
  2008.

\bibitem[De~Santis et~al.(2022)De~Santis, Goeman, Hemerik, and
  Finos]{desantis2022inference}
R.~De~Santis, J.~J. Goeman, J.~Hemerik, and L.~Finos.
\newblock Inference in generalized linear models with robustness to
  misspecified variances, 2022.
\newblock URL \url{https://arxiv.org/abs/2209.13918}.

\bibitem[Debeer and Strobl(2020)]{debeer2020conditional}
D.~Debeer and C.~Strobl.
\newblock Conditional permutation importance revisited.
\newblock \emph{BMC bioinformatics}, 21\penalty0 (1):\penalty0 1--30, 2020.

\bibitem[Dobriban(2022)]{dobriban2021consistency}
E.~Dobriban.
\newblock {Consistency of invariance-based randomization tests}.
\newblock \emph{The Annals of Statistics}, 50\penalty0 (4):\penalty0 2443 --
  2466, 2022.
\newblock \doi{10.1214/22-AOS2200}.
\newblock URL \url{https://doi.org/10.1214/22-AOS2200}.

\bibitem[Dwass(1957)]{dwass1957}
M.~Dwass.
\newblock Modified randomization tests for nonparametric hypotheses.
\newblock \emph{The Annals of Mathematical Statistics}, 28:\penalty0 181--187,
  1957.

\bibitem[Eaton(1989)]{eaton1989group}
M.~L. Eaton.
\newblock Group invariance applications in statistics.
\newblock In \emph{Regional conference series in Probability and Statistics},
  pages i--133. JSTOR, 1989.

\bibitem[Eden and Yates(1933)]{eden1933validity}
T.~Eden and F.~Yates.
\newblock On the validity of fisher's z test when applied to an actual example
  of non-normal data.(with five text-figures.).
\newblock \emph{The Journal of Agricultural Science}, 23\penalty0 (1):\penalty0
  6--17, 1933.

\bibitem[Efron(1969)]{efron1969student}
B.~Efron.
\newblock Student's t-test under symmetry conditions.
\newblock \emph{Journal of the American Statistical Association}, 64\penalty0
  (328):\penalty0 1278--1302, 1969.
\newblock ISSN 01621459.
\newblock URL \url{http://www.jstor.org/stable/2286068}.

\bibitem[Eklund et~al.(2016)Eklund, Nichols, and Knutsson]{eklund2016cluster}
A.~Eklund, T.~E. Nichols, and H.~Knutsson.
\newblock Cluster failure: Why fmri inferences for spatial extent have inflated
  false-positive rates.
\newblock \emph{Proceedings of the national academy of sciences}, 113\penalty0
  (28):\penalty0 7900--7905, 2016.

\bibitem[Fisher(1915)]{fisher1915frequency}
R.~A. Fisher.
\newblock Frequency distribution of the values of the correlation coefficient
  in samples from an indefinitely large population.
\newblock \emph{Biometrika}, 10\penalty0 (4):\penalty0 507--521, 1915.

\bibitem[Fisher(1925)]{fisher1925applications}
R.~A. Fisher.
\newblock Applications of "student's" distribution.
\newblock \emph{Metron}, 5:\penalty0 90--104, 1925.
\newblock URL \url{https://hdl.handle.net/2440/15187}.

\bibitem[Fisher(1935)]{fisher1935}
R.~A. Fisher.
\newblock \emph{The design of experiments}.
\newblock Oliver and Boyd, 1935.

\bibitem[Gao et~al.(2010)Gao, Becker, Becker, Starmer, and
  Province]{gao2010avoiding}
X.~Gao, L.~C. Becker, D.~M. Becker, J.~D. Starmer, and M.~A. Province.
\newblock Avoiding the high bonferroni penalty in genome-wide association
  studies.
\newblock \emph{Genetic Epidemiology: The Official Publication of the
  International Genetic Epidemiology Society}, 34\penalty0 (1):\penalty0
  100--105, 2010.

\bibitem[Girardi et~al.(2022)Girardi, Vesely, Lakens, Alto{\`e}, Pastore,
  Calcagn{\`\i}, and Finos]{girardi2022post}
P.~Girardi, A.~Vesely, D.~Lakens, G.~Alto{\`e}, M.~Pastore, A.~Calcagn{\`\i},
  and L.~Finos.
\newblock Post-selection inference in multiverse analysis (pima): an
  inferential framework based on the sign flipping score test.
\newblock \emph{arXiv preprint arXiv:2210.02794}, 2022.

\bibitem[Good(2005)]{good2005permutation}
P.~Good.
\newblock \emph{Permutation, Parametric, and Bootstrap Tests of Hypotheses (3rd
  ed.)}.
\newblock Springer-Verlag, New York, 2005.

\bibitem[Hemerik and Goeman(2018{\natexlab{a}})]{Hemerik_confSAM_2018}
J.~Hemerik and J.~Goeman.
\newblock {confSAM}.
\newblock \url{https://cran.r-project.org/web/packages/confSAM/index.html},
  2018{\natexlab{a}}.

\bibitem[Hemerik and Goeman(2018{\natexlab{b}})]{hemerik2018exact}
J.~Hemerik and J.~J. Goeman.
\newblock Exact testing with random permutations.
\newblock \emph{TEST}, 27\penalty0 (4):\penalty0 811--825, 2018{\natexlab{b}}.

\bibitem[Hemerik and Goeman(2018{\natexlab{c}})]{hemerik2018false}
J.~Hemerik and J.~J. Goeman.
\newblock False discovery proportion estimation by permutations: confidence for
  significance analysis of microarrays.
\newblock \emph{Journal of the Royal Statistical Society: Series B (Statistical
  Methodology)}, 80\penalty0 (1):\penalty0 137--155, 2018{\natexlab{c}}.

\bibitem[Hemerik and Goeman(2021)]{hemerik2021another}
J.~Hemerik and J.~J. Goeman.
\newblock Another look at the lady tasting tea and differences between
  permutation tests and randomisation tests.
\newblock \emph{International Statistical Review}, 89\penalty0 (2):\penalty0
  367--381, 2021.

\bibitem[Hemerik et~al.(2019)Hemerik, Solari, and
  Goeman]{hemerik2019permutation}
J.~Hemerik, A.~Solari, and J.~Goeman.
\newblock Permutation-based simultaneous confidence bounds for the false
  discovery proportion.
\newblock \emph{Biometrika}, 106\penalty0 (3):\penalty0 635--649, 2019.

\bibitem[Hemerik et~al.(2020)Hemerik, Goeman, and Finos]{hemerik2020robust}
J.~Hemerik, J.~J. Goeman, and L.~Finos.
\newblock Robust testing in generalized linear models by sign flipping score
  contributions.
\newblock \emph{Journal of the Royal Statistical Society: Series B (Statistical
  Methodology)}, 82\penalty0 (3):\penalty0 841--864, 2020.

\bibitem[Hemerik et~al.(2021)Hemerik, Thoresen, and
  Finos]{hemerik2021permutation}
J.~Hemerik, M.~Thoresen, and L.~Finos.
\newblock Permutation testing in high-dimensional linear models: an empirical
  investigation.
\newblock \emph{Journal of Statistical Computation and Simulation}, 91\penalty0
  (5):\penalty0 897--914, 2021.

\bibitem[Hotelling(1953)]{hotelling1953new}
H.~Hotelling.
\newblock New light on the correlation coefficient and its transforms.
\newblock \emph{Journal of the Royal Statistical Society. Series B
  (Methodological)}, 15\penalty0 (2):\penalty0 193--232, 1953.

\bibitem[Kofler and Schl{\"o}tterer(2012)]{kofler2012gowinda}
R.~Kofler and C.~Schl{\"o}tterer.
\newblock Gowinda: unbiased analysis of gene set enrichment for genome-wide
  association studies.
\newblock \emph{Bioinformatics}, 28\penalty0 (15):\penalty0 2084--2085, 2012.

\bibitem[Koning(2022{\natexlab{a}})]{Koning_Database_of_Near_2022}
N.~Koning.
\newblock {Database of Near Oracle Subgroups}.
\newblock \url{https://github.com/nickwkoning/NOSdata}, 2022{\natexlab{a}}.

\bibitem[Koning(2022{\natexlab{b}})]{Koning_Near_Oracle_Subgroups_2022}
N.~Koning.
\newblock {Near Oracle Subgroups}.
\newblock \url{https://github.com/nickwkoning/NOS}, 2022{\natexlab{b}}.

\bibitem[Koning(2022{\natexlab{c}})]{Koning_fast_confSAM_2022}
N.~Koning.
\newblock {fast confSAM}.
\newblock \url{https://github.com/nickwkoning/fastconfSAM}, 2022{\natexlab{c}}.

\bibitem[Langsrud(2005)]{langsrud2005rotation}
{\O}.~Langsrud.
\newblock Rotation tests.
\newblock \emph{Statistics and computing}, 15\penalty0 (1):\penalty0 53--60,
  2005.

\bibitem[Lehmann and Romano(2005)]{lehmann2005testing}
E.~L. Lehmann and J.~P. Romano.
\newblock \emph{Testing statistical hypotheses}.
\newblock Springer Science \& Business Media, 2005.

\bibitem[Li and Tibshirani(2013)]{li2013finding}
J.~Li and R.~Tibshirani.
\newblock Finding consistent patterns: a nonparametric approach for identifying
  differential expression in rna-seq data.
\newblock \emph{Statistical methods in medical research}, 22\penalty0
  (5):\penalty0 519--536, 2013.

\bibitem[Meinshausen(2006)]{meinshausen2006false}
N.~Meinshausen.
\newblock False discovery control for multiple tests of association under
  general dependence.
\newblock \emph{Scandinavian Journal of Statistics}, 33\penalty0 (2):\penalty0
  227--237, 2006.

\bibitem[Meinshausen et~al.(2011)Meinshausen, Maathuis, B{\"u}hlmann,
  et~al.]{meinshausen2011asymptotic}
N.~Meinshausen, M.~H. Maathuis, P.~B{\"u}hlmann, et~al.
\newblock Asymptotic optimality of the westfall--young permutation procedure
  for multiple testing under dependence.
\newblock \emph{The Annals of Statistics}, 39\penalty0 (6):\penalty0
  3369--3391, 2011.

\bibitem[{\acroauthor{OEIS Foundation
  Inc.}{OEIS}}(2021{\natexlab{a}})]{oeisA006116}
{\acroauthor{OEIS Foundation Inc.}{OEIS}}.
\newblock The on-line encyclopedia of integer sequences.
\newblock \url{http://oeis.org/A006116}, 2021{\natexlab{a}}.

\bibitem[{\acroauthor{OEIS Foundation
  Inc.}{OEIS}}(2021{\natexlab{b}})]{oeisA022166}
{\acroauthor{OEIS Foundation Inc.}{OEIS}}.
\newblock The on-line encyclopedia of integer sequences.
\newblock \url{http://oeis.org/A022166}, 2021{\natexlab{b}}.

\bibitem[{\acroauthor{OEIS Foundation
  Inc.}{OEIS}}(2021{\natexlab{c}})]{oeisA076766}
{\acroauthor{OEIS Foundation Inc.}{OEIS}}.
\newblock The on-line encyclopedia of integer sequences.
\newblock \url{http://oeis.org/A076766}, 2021{\natexlab{c}}.

\bibitem[{\acroauthor{OEIS Foundation
  Inc.}{OEIS}}(2021{\natexlab{d}})]{oeisA076831}
{\acroauthor{OEIS Foundation Inc.}{OEIS}}.
\newblock The on-line encyclopedia of integer sequences.
\newblock \url{http://oeis.org/A076831}, 2021{\natexlab{d}}.

\bibitem[Onghena(2018)]{onghena2018randomization}
P.~Onghena.
\newblock Randomization tests or permutation tests? {A} historical and
  terminological clarification.
\newblock \emph{Randomization, masking, and allocation concealment}, pages
  209--227, 2018.

\bibitem[Pesarin and Salmaso(2010)]{pesarin2010permutation}
F.~Pesarin and L.~Salmaso.
\newblock \emph{Permutation tests for complex data: theory, applications and
  software}.
\newblock John Wiley \& Sons, 2010.

\bibitem[Phipson and Smyth(2010)]{phipson2010permutation}
B.~Phipson and G.~K. Smyth.
\newblock Permutation p-values should never be zero: calculating exact p-values
  when permutations are randomly drawn.
\newblock \emph{Statistical applications in genetics and molecular biology},
  9\penalty0 (1):\penalty0 39, 2010.

\bibitem[Skitovitch(1953)]{skitovitch1953property}
V.~P. Skitovitch.
\newblock On a property of the normal distribution.
\newblock \emph{Doklady Akad. Nauk SSSR (N.S)}, 89:\penalty0 217--219, 1953.

\bibitem[Slepian(1968)]{slepian1968group}
D.~Slepian.
\newblock Group codes for the gaussian channel.
\newblock \emph{Bell System Technical Journal}, 47\penalty0 (4):\penalty0
  575--602, 1968.

\bibitem[Sloane et~al.(1996)Sloane, Hardin, Smith, et~al.]{sloane1996spherical}
N.~J.~A. Sloane, R.~Hardin, W.~Smith, et~al.
\newblock Tables of spherical codes.
\newblock \url{http://neilsloane.com/packings/}, 1996.
\newblock Accessed: 2021-11-19.

\bibitem[Smeets et~al.(2013)Smeets, Kroese, Evers, and
  de~Ridder]{smeets2013allured}
P.~A. Smeets, F.~M. Kroese, C.~Evers, and D.~T. de~Ridder.
\newblock Allured or alarmed: counteractive control responses to food
  temptations in the brain.
\newblock \emph{Behavioural brain research}, 248:\penalty0 41--45, 2013.

\bibitem[Solari et~al.(2014)Solari, Finos, and Goeman]{solari2014rotation}
A.~Solari, L.~Finos, and J.~J. Goeman.
\newblock Rotation-based multiple testing in the multivariate linear model.
\newblock \emph{Biometrics}, 70\penalty0 (4):\penalty0 954--961, 2014.

\bibitem[Southworth et~al.(2009)Southworth, Kim, and
  Owen]{southworth2009properties}
L.~K. Southworth, S.~K. Kim, and A.~B. Owen.
\newblock Properties of balanced permutations.
\newblock \emph{Journal of Computational Biology}, 16\penalty0 (4):\penalty0
  625--638, 2009.

\bibitem[Tusher et~al.(2001)Tusher, Tibshirani, and
  Chu]{tusher2001significance}
V.~G. Tusher, R.~Tibshirani, and G.~Chu.
\newblock Significance analysis of microarrays applied to the ionizing
  radiation response.
\newblock \emph{Proceedings of the National Academy of Sciences}, 98\penalty0
  (9):\penalty0 5116--5121, 2001.

\bibitem[Vesely et~al.(2021)Vesely, Finos, and Goeman]{vesely2021permutation}
A.~Vesely, L.~Finos, and J.~J. Goeman.
\newblock Permutation-based true discovery guarantee by sum tests.
\newblock \emph{arXiv preprint arXiv:2102.11759}, 2021.

\bibitem[Westfall and Troendle(2008)]{westfall2008multiple}
P.~H. Westfall and J.~F. Troendle.
\newblock Multiple testing with minimal assumptions.
\newblock \emph{Biometrical Journal: Journal of Mathematical Methods in
  Biosciences}, 50\penalty0 (5):\penalty0 745--755, 2008.

\bibitem[Westfall and Young(1993)]{westfall1993resampling}
P.~H. Westfall and S.~S. Young.
\newblock \emph{Resampling-based multiple testing: Examples and methods for
  p-value adjustment}, volume 279.
\newblock John Wiley \& Sons, 1993.

\bibitem[Winkler et~al.(2014)Winkler, Ridgway, Webster, Smith, and
  Nichols]{winkler2014permutation}
A.~M. Winkler, G.~R. Ridgway, M.~A. Webster, S.~M. Smith, and T.~E. Nichols.
\newblock Permutation inference for the general linear model.
\newblock \emph{Neuroimage}, 92:\penalty0 381--397, 2014.

\bibitem[Winkler et~al.(2016)Winkler, Ridgway, Douaud, Nichols, and
  Smith]{winkler2016faster}
A.~M. Winkler, G.~R. Ridgway, G.~Douaud, T.~E. Nichols, and S.~M. Smith.
\newblock Faster permutation inference in brain imaging.
\newblock \emph{NeuroImage}, 141:\penalty0 502--516, 2016.

\bibitem[Young(2019)]{young2019channeling}
A.~Young.
\newblock Channeling fisher: Randomization tests and the statistical
  insignificance of seemingly significant experimental results.
\newblock \emph{The Quarterly Journal of Economics}, 134\penalty0 (2):\penalty0
  557--598, 2019.

\end{thebibliography}

\end{document}